\documentclass[aps,prd,twocolumn,nofootinbib,superscriptaddress]{revtex4-2}
\usepackage{graphicx}
\usepackage{dcolumn}
\usepackage{mathtools}
\usepackage{bm}
\usepackage[dvipsnames]{xcolor}
\usepackage{algpseudocode}
\usepackage{tcolorbox}
\usepackage{siunitx}
\usepackage{caption}
\usepackage{subcaption}
\usepackage{amssymb,amsmath,amsfonts,amsthm}
\usepackage{IEEEtrantools}
\usepackage{appendix} 
\usepackage[pagebackref=false,
  hyperindex=true,
  citecolor=blue,
  colorlinks=true,
  linkcolor=blue,
urlcolor=blue]{hyperref}
\usepackage[capitalize]{cleveref}
\newcommand{\Zeta}{\mathrm{Z}}

\let\originalleft\left
\let\originalright\right
\renewcommand{\left}{\mathopen{}\mathclose\bgroup\originalleft}
\renewcommand{\right}{\aftergroup\egroup\originalright}

\definecolor{blue}{rgb}{0.36, 0.54, 0.85}
\definecolor{amaranth}{rgb}{0.9, 0.17, 0.31}
\definecolor{pink}{rgb}{0.87, 0.56, 0.81}
\definecolor{ao}{rgb}{0.0, 0.5, 0.0}
\definecolor{maroon}{rgb}{0.76, 0.13, 0.28}
\definecolor{cardinal}{rgb}{0.77, 0.12, 0.23}
\definecolor{lightcardinal}{rgb}{0.97, 0.42, 0.53}
\definecolor{frenchlila}{rgb}{0.53, 0.38, 0.56}
\definecolor{yellow}{rgb}{1.0, 1.0, 0.87}
\definecolor{lightseagreen}{rgb}{0.45, 0.85, 0.58}
\definecolor{gray}{rgb}{0.9, 0.9, 0.9}
\definecolor{lightblue}{rgb}{0.66, 0.84, 0.96}

\begin{document}

\title{Contamination of transient gravitational waves in LISA data by
gaps and glitches}

\author{Jonathan R. Gair}
\email{jonathan.gair@aei.mpg.de}
\affiliation{Max-Planck-Institut f\"ur Gravitationsphysik,
Albert-Einstein-Institut,  Am M\"uhlenberg 1, 14476 Potsdam-Golm, Germany}

\author{Senwen Deng}
\email{deng@apc.in2p3.fr}
\affiliation{Universit\'e Paris Cit\'e, CNRS, Astroparticule et
Cosmologie, F-75013 Paris, France}

\author{Stanislav Babak}
\email{stas@apc.in2p3.fr}
\affiliation{Universit\'e Paris Cit\'e, CNRS, Astroparticule et
Cosmologie, F-75013 Paris, France}

\date{\today}

\begin{abstract}
  We present a probabilistic framework to quantify the impact of
  artefacts (glitches and gaps) in LISA data on transient
  gravitational wave signals. By modeling both artefacts and
  transient signals as independent Poisson processes, and
  characterising the contaminating effect of an artefact by an
  associated dead time, we estimate the probability distribution of
  the total contamination time during the observation period using a
  Normal approximation under various contamination scenarios. Using
  the same approach, we also estimate the probability that a
  population of transient signals contaminates each other. We
  demonstrate the validity of the Normal approximation by comparing
  it to the numerical distribution obtained via simulations. Our
  approach provides a rapid means to assess the potential impact of
  glitches and gaps on LISA science, and can be used as a figure of
  merit to evaluate different instrumental scenarios.
\end{abstract}

\maketitle

\section{\label{sec:intro}Introduction}

The Laser Interferometer Space Antenna (LISA) is a planned space-borne
gravitational wave (GW) observatory
that will be launched around 2035. It will consist of three
spacecrafts in heliocentric orbits that form the corners of an
approximately equilateral triangle, each shielding two test masses in
free-fall. The spacecrafts will exchange laser light, allowing the
proper distance between the distant test masses to be measured using
transponding interferometry.
The optical path is affected by passing GWs
manifesting themselves as a change of the laser frequency. Regularly
during the duration of the mission, the spacecraft will undergo
maintenance and approximately every two weeks the antennae will be
redirected. These operations will lead to a significant disturbance
of the spacecrafts and most likely cause periods of bad data quality:
these periods of time we treat as scheduled gaps in the data. In
addition, we might expect to have unintended gaps in the data caused
by, for example, (i) failures and necessity to catch-and-release the
test mass; or (ii) minutes-long dropouts in the transmitted data.  In
addition to the gaps, we also expect ``glitches" in the data caused
by instrumental and environmental disturbances
\cite{armano_sub_femto_2016}.
Glitches are transient features in the instrumental noise, and the
LISA Pathfinder mission identified at least three categories of glitches
\cite{lisa_pathfinder_collaboration_transient_2022}:
(i) impulse-carrying glitches; (ii) fast low-impulse glitches; and
(iii) micrometeoroid impacts, which are well understood and can be
suppressed by calibration. The scientific return of LISA will depend
on the integrated observation time that is free of artefacts. The
nominal mission duration of 4.5 years, with an anticipated duty cycle
of more than 82\%, is expected to provide a sufficiently long
observation time to
meet the scientific objectives of LISA
\cite{colpi_lisa_2024}.



LISA data will be signal-dominated. We can broadly separate the
anticipated GW signals into two groups: long-lived and transient
signals.  The characteristic time scale separating these categories
is given by the timescale over which LISA's position changes due to
its orbital motion, which is about a month. Long-lived GWs are those
with duration significantly longer than a few months. Long-lived
sources gradually accumulate signal-to-noise ratio (SNR) over a long
period of time (longer than 1-2 months). Examples of long-lived GW
sources in the LISA band include: a) Galactic white dwarf binaries,
which are mildly relativistic sources emitting almost
monochromatic signal over the whole duration of the LISA mission; b)
Stellar mass
black hole binaries, which are the primary sources that are observed
merging in the band of the ground-based detectors by the
LIGO-VIRGO-Kagra collaboration \cite{collaboration_prospects_2020}.
LISA will observe between a few and a few dozen of these binaries
during the early inspiral stage. Stellar mass binaries at high
frequency (above 20 mHz), if sufficiently massive, will inspiral out
of the LISA band and will merge several years later in the band of
third generation of ground-based detectors like Cosmic Explorer
\cite{dwyer_gravitational_2015,abbott_exploring_2017}
and the Einstein Telescope
\cite{hild_pushing_2008,hild_sensitivity_2011,punturo_einstein_2010};
and c) EMRIs (extreme mass ratio inspirals) -- systems that result
from the capture of a stellar mass black hole by a massive black hole
(MBH) in dense galactic nuclei with a subsequent inspiral in the
vicinity of an MBH. The signal could last months to years before the
stellar-mass black hole plunges into the horizon.

Among transient sources the most anticipated are merging massive
black hole binaries (MBHBs). Although the signal from those binaries
could be present in the data for a long time, only the late inspiral,
merger and ring-down parts of the signal are detectable. Depending on
the mass of MBHBs and the strength of the signal, we will observe
MBHBs for periods of between a few hours and a month. The immediate
pre-merger and the merger parts of the signal are the
most important for the parameter estimation as they dominate the
signal-to-noise ratio (SNR) and during this phase the binary is fully
relativistic (orbital speed is a significant fraction of the speed of light).

Gaps and glitches contaminate both types of sources, but will have
different importance in the two cases. In the case of long-lived
sources, the presence of gaps leads to a loss of SNR and could bias
estimated parameters if gaps are not handled properly in the
template. Mitigation procedures include inpainting and signal
reconstruction in the gap
\cite{baghi_gravitational_wave_2019,blelly_sparse_2022} or handling
the gap using non-diagonal likelihoods~\cite{Burke:2025bun}.
Glitches can be reduced to gaps by simply
gating the data around the glitch. Alternatively, glitches can be
considered as ``unmodelled" signals and fitted together with GWs, see,
for example,
\textcite{robson_detecting_2019,pankow_mitigation_2018,Muratore:2025knh}.
Since the gaps are short
compared to the duration of long-lived signals, they do not carry a lot of
weight and the techniques suggested above should work well. Glitches
and gaps are much more dangerous if they occur next to the merger of
MBHBs. In this case, we might lose a significant fraction of SNR and
vital information for proper parameter estimation.


In this work, we investigate the possibility that artefacts (gaps
and/or glitches) contaminate the transient signals in LISA using a
probabilistic approach. We simplify the problem by considering the
appearance of astrophysical transient sources (bursts, MBHB mergers)
as a Poisson process. The second Poisson process describes the
occurrence of gaps (or glitches) in the data. With each gap (glitch)
we associate a ``dead time'', meaning that any astrophysical event
that falls into this time interval is lost. This is a crude
approximation and gives a conservative estimate of the astrophysical
events lost to the artefacts (as we collectively call gaps and
glitches). The dead time associated with an artefact is a combination of
the typical duration of an artefact and the time interval relative to
the merger within which the GW signal would be lost or severely
contaminated. We build a mathematical (semi-)analytical model of the
data and investigate different scenarios of contamination of the
astrophysical transients by artefacts. The objective of this work is
to develop quick ways to assess the potential impact of glitches and
gaps on LISA science under varying instrumental scenarios, which can
be used as a Figure of Merit for evaluating different instrumental
scenarios as the mission is developed.

The paper is organized as follows. In \cref{sec:methods}, we present
the mathematical formulation of the problem and propose an
approximation based on the normal distribution to evaluate the time
of contamination (or self-contamination). In \cref{sec:results}, we
compare the normal approximation with the numerical distribution
obtained from simulations to demonstrate the validity of the
approximation. In \cref{sec:constraints}, we provide an example of
how to constrain the parameters characterizing the contamination
process. Finally, we give a summary in \cref{sec:summary}.


\section{\label{sec:methods}Methods}
We model the observations of populations of artefacts and
astrophysical transient signals as two independent Poisson processes
with the corresponding rates $\lambda$ and $\mu$.
As mentioned above, the contamination of GW signals by artefacts is
encoded in the ``dead time" attached to the occurrence of
each artefact event, the overlapping dead times due to proximity of
artefacts combine into
a common contamination interval. When the artefact is a gap, the
dead time is equal to the duration of the
gap. In the case of a glitch, the dead time combines together the
duration (ringing) of a glitch and its proximity to the merger (or a
characteristic time of arrival of a GW burst).
We assume that each GW signal falling within the dead time of an
artefact is lost.
In this work, we consider three distributions of dead times:
constant, uniform distribution, and an exponential
distribution. We will specify which distribution is used, and to
each artefact we will attach a dead time drawn from the specific distribution.


It is possible that an artefact occurs within the dead time of the
previous artefact. In this case, we use two possible
treatments: (i) \emph{merged scenario}; and 
(ii) \emph{reset scenario}.
In the first (merged) scenario, the overlapping dead times
are combined
in union to form a contamination
interval. This represents a scenario in which
gaps and glitches do not affect each other (for example,
glitches of different nature arising in different parts of the
instrument) but both contaminate the data.
The merged contamination interval is at least as long as the longest
dead time, up to the sum of all dead times.
In the second (reset) scenario, we
assume that the contamination is
reset whenever a new artefact event occurs. That means the
contamination of the previous artefact terminates at the time when
the succeeding
artefact occurs, and the end of the contamination interval is
defined by the occurrence time and the dead time of the last
artefact in the dead-time overlapping sequence. The length of the
contamination interval is then given by the time separation between the
first and the last artefacts plus the dead time of the last artefact.
This represents a
scenario in which each glitch is strongest at its onset and then
decays. We assume each glitch represents a sufficiently strong
perturbation to the instrument that it erases memory of the prior
state of the system. The end of the contamination period is then
determined by the length of the last glitches' dead time.
The reset scenario could be applicable to glitches of the same
physical nature, for example, two consecutive micrometeoroids
impacting the same spacecraft.
We note that both scenarios are
equivalent if the dead time is constant for all artefacts.

Note that we are representing a GW event
occurrence by a single time, for example the merger. Real GW signals
are not point-like. However, the duration of the signal can be
absorbed into the definition of the artefact dead time, such that
it represents the range of times at which the merger of a signal
could fall and suffer significant contamination.
We use
\(T_\text{ctmn}\) to denote the total contamination time summed over
all artefacts that occur during the observation period \(T_\text{obs}\).
The number of GW signals lost to contamination then has the
probability mass function (PMF)
\begin{IEEEeqnarray}{rCl}
P(k|T_\text{ctmn}) &=& \sum_{l=k}^\infty \frac{e^{-\mu
T_\text{ctmn}} (\mu T_\text{ctmn})^l}{l!} \binom{l}{k}
\nonumber\\ && \hphantom{\sum_{l=k}^\infty} \times \left(
\frac{T_\text{ctmn}}{T_\text{obs}} \right)^{\!k}
\left( 1 - \frac{T_\text{ctmn}}{T_\text{obs}} \right)^{\!l-k},
\end{IEEEeqnarray}
where we use $\mu$ to denote the Poisson rate of MBHB mergers. The
marginalized PMF for the number of GW signals contaminated during the
observation time is then given by
\begin{equation}
P(k) = \int_0^{T_\text{obs}} P(k|T_\text{ctmn}) \pi(T_\text{ctmn})
\, \mathrm{d} T_\text{ctmn},
\label{eq:unconditional_ctmn_num_pmf}
\end{equation}
where \(\pi(T_\text{ctmn})\) is the probability density function
(PDF) for the distribution of
\(T_\text{ctmn}\).

We can use the sane formalism to address the question of
self-contamination of GW
signals, i.e., contamination arising from signals within the same
population. In this case, ``contamination'' means that two signals
overlap sufficiently in time (e.g., two merging MBHBs occur close to
each other in
time) that they must be treated jointly.  The difference with the
previous model is that both events belong to the same Poisson process
with rate \(\mu\).
We can still assess the probability \(P(k)\), of having two (or more)
overlapping signals during the observation, and we will consider this
scenario in
\cref{sec:self-contamination}.


\subsection{Normal approximation}\label{sec:normal_approx}
The data acquired during the LISA Pathfinder operation contain
several population of glitches
\cite{lisa_pathfinder_collaboration_transient_2022}.
Extrapolating these
results to LISA, we
expect that about 15-20\% of the data might be contaminated by
artefacts, implying that the glitches and gaps are numerous (given
their usually short duration).
Based on the law of large numbers, we expect to be able to approximate
\(\pi(T_\text{ctmn})\) well by a Normal distribution. In this
subsection, we show how to construct a mathematical model for
different scenarios (merged, reset) and based on the assumption of a
Gaussian approximation. In \cref{sec:results}, we will verify the
validity of this
approximation and assess its limitations.

We start with a general description and then apply it to particular
cases. Let us introduce random independent variables \(\Delta T_i\)
and \(\delta t_i\), where \(\Delta T_i\) is the duration of the
\(i\)'th contamination interval (merged or reset combination of dead
times) and \(\delta t_i\) is the time elapsed between the end of the
\(i\)'th and the start of the \((i+1)\)'th contamination intervals. We
can construct an associated random variable
\begin{equation}
f_n = \frac{ \frac{1}{n} \sum_{i=1}^{n} \Delta T_i }
{\frac{1}{n} \sum_{i=1}^{n} (\Delta T_i + \delta t_i)},
\label{eq:f_n_general}
\end{equation}
which represents the fraction of time contaminated by glitches during
the observation period which contains the first \(n\) contamination intervals.
We assume (and verify it later) that \(f_n\)  can be well
approximated by a Normal distribution, \(f_n \sim N
\left(\mathbb{E}\left[f_n\right],
\mbox{Var}\left[f_n\right]\right)\),
in which case the total contamination
time \(T_\text{ctmn}\) also follows a Normal distribution:
\(T_\text{ctmn} \sim N \left(T_\text{obs}\mathbb{E}\left[f_n\right],
T_\text{obs}^2 \mbox{Var}\left[f_n\right]\right)\).

We need to compute the expectation and variance of \(f_n\).
Let us consider two random variables \(X\) and \(Y\) which are the
average of the sum of \(n\)
independent and identically distributed random variables, denoted by
\(X_i\) and \(Y_i\) respectively.
It follows that \(\mathbb{E}\left[X\right] =
\mathbb{E}\left[X_i\right]\), \(\mathbb{E}\left[Y\right] =
\mathbb{E}\left[Y_i\right]\),
\(\mbox{Var}\left[X\right] = \frac{1}{n} \mbox{Var}\left[X_i\right]\) and
\(\mbox{Var}\left[Y\right] = \frac{1}{n}
\mbox{Var}\left[Y_i\right]\). Moreover, whenever pairs \((X_i, Y_i)\) and
\((X_j, Y_j)\)
are independent for \(i \neq j\), we have \(\mbox{Cov}\left[X,
Y\right] = \frac{1}{n} \mbox{Cov}\left[X_i, Y_i\right]\).

The expectation and variance of \(\frac{X}{Y}\) can be computed with
an expansion in \(1/n\) (where $n$ is the size of the ensemble),
\begin{IEEEeqnarray}{rCl}
\mathbb{E}\left[\frac{X}{Y}\right] &=&
\frac{\mathbb{E}\left[X\right]}{\mathbb{E}\left[Y\right]}
+ \frac{\mathbb{E}\left[X\right]}{\mathbb{E}\left[Y\right]^3}
\mbox{Var}\left[Y\right]
\nonumber\\ && - \frac{1}{\mathbb{E}\left[Y\right]^2}
\mbox{Cov}\left[X, Y\right]
+ o(1/n)
\\ &=& \frac{\mathbb{E}\left[X\right]}{\mathbb{E}\left[Y\right]} +
O(1/n), \label{eq:mean_x_y}
\\ \mbox{Var}\left[\frac{X}{Y}\right] &=&
\frac{1}{\mathbb{E}\left[Y\right]^2} \mbox{Var}\left[X\right]
+ \frac{\mathbb{E}\left[X\right]^2}{\mathbb{E}\left[Y\right]^4}
\mbox{Var}\left[Y\right]
\nonumber\\ && -
\frac{2\mathbb{E}\left[X\right]}{\mathbb{E}\left[Y\right]^3}
\mbox{Cov}\left[X, Y\right]
+ o(1/n). \label{eq:variance_x_y}
\end{IEEEeqnarray}
Applying the above formulae to the random variable \(f_n\) defined in
\cref{eq:f_n_general} we get
\begin{IEEEeqnarray}{rCl}
\mathbb{E}\left[f_n\right] &\approx& \frac{\mathbb{E}\left[\Delta T_i\right]}
{\mathbb{E}\left[\Delta T_i\right]+\mathbb{E}\left[\delta t_i\right]},
\label{eq:mean_fn_general}
\\ \mbox{Var}\left[f_n\right] &\approx&
\frac{1}{n}\left(\vphantom{\frac{ \mbox{Var}\left[\Delta T_i\right] }
{ \left(\mathbb{E}\left[\Delta
T_i\right]+\mathbb{E}\left[\delta t_i\right]\right)^2 }
+ \frac{\mathbb{E}\left[\Delta T_i\right]^2}
{ \left(\mathbb{E}\left[\Delta
T_i\right]+\mathbb{E}\left[\delta t_i\right]\right)^4 }
\left(\mbox{Var}\left[\Delta T_i\right] +
\mbox{Var}\left[\delta t_i\right] \right)
- \frac{2\mathbb{E}\left[\Delta T_i\right]}{
  \left(\mathbb{E}\left[\Delta T_i\right]
+\mathbb{E}\left[\delta t_i\right]\right)^3 }
\mbox{Var}\left[\Delta T_i\right]}\right. \frac{
\mbox{Var}\left[\Delta T_i\right] }
{ \left(\mathbb{E}\left[\Delta T_i\right]+\mathbb{E}\left[\delta
t_i\right]\right)^2 }
\nonumber\\ && \hphantom{\frac{1}{n}} +
\frac{\mathbb{E}\left[\Delta T_i\right]^2}
{ \left(\mathbb{E}\left[\Delta T_i\right]+\mathbb{E}\left[\delta
t_i\right]\right)^4 }
\left(\mbox{Var}\left[\Delta T_i\right] + \mbox{Var}\left[\delta
t_i\right] \right)
\nonumber\\ && \hphantom{\frac{1}{n}} -
\frac{2\mathbb{E}\left[\Delta T_i\right]}{
\left(\mathbb{E}\left[\Delta T_i\right]+\mathbb{E}\left[\delta
t_i\right]\right)^3 }
\mbox{Var}\left[\Delta T_i\right] \left.\vphantom{\frac{
\mbox{Var}\left[\Delta T_i\right] }
{ \left(\mathbb{E}\left[\Delta
T_i\right]+\mathbb{E}\left[\delta t_i\right]\right)^2 }
+ \frac{\mathbb{E}\left[\Delta T_i\right]^2}
{ \left(\mathbb{E}\left[\Delta
T_i\right]+\mathbb{E}\left[\delta t_i\right]\right)^4 }
\left(\mbox{Var}\left[\Delta T_i\right] +
\mbox{Var}\left[\delta t_i\right] \right)
- \frac{2\mathbb{E}\left[\Delta T_i\right]}{
  \left(\mathbb{E}\left[\Delta T_i\right]
+\mathbb{E}\left[\delta t_i\right]\right)^3 }
\mbox{Var}\left[\Delta T_i\right]}\right).
\label{eq:var_fn_general}
\end{IEEEeqnarray}

The distribution of \(\delta t_i\) is exponential with rate \(\lambda\), so that
\begin{align}
\mathbb{E}\left[\delta t_i\right] & = \frac{1}{\lambda},
\label{eq:mean_delta_ti}       \\
\mbox{Var}\left[\delta t_i\right] & = \frac{1}{\lambda^2}.
\label{eq:variance_delta_ti}
\end{align}
These are general expressions and we will further develop them for
the merged and reset
scenarios in
\cref{sec:merged_ctmn_interval_distribution,sec:reset_ctmn_interval_distribution}
respectively.

For the Normal approximation to perform well, it is
necessary that the number of contamination intervals is large enough.
If there are many long contamination periods, then it is likely that
they will merge to form a few long contamination intervals, and the
Normal approximation would break down.

\subsection{Constant dead time scenario}\label{sec:fixed_lost_period}

Before going to more complicated scenarios, we consider the
simple case where the dead time attached to every
glitch is constant \(\tau\).
This simplifies the general description given in the previous subsection.
We start with defining the nonoverlapped dead time $\Delta T_{g,i}$
attached to the \(i\)'th artefact, note that it is equal to $\tau$ if
the next artefact falls outside the constant dead time, otherwise
$\Delta T_{g,i} < \tau$. We also introduce the time interval between
the \(i\)'th and \((i+1)\)'th artefacts, as $\delta t_{g,i}$, it is equal
to $\Delta T_{g,i}$ if the \((i+1)\)'th artefact falls into the dead
time of the \((i)\)'th artefact. Note that this is the time
difference between the start of one contamination period and the
start of the next and so it is not the same as $\delta t_i$ introduced above.


The variables \(\Delta T_{g,i}\) and \(\delta t_{g,i}\) are correlated but
\((\Delta T_{g,i}, \delta t_{g,i})\) and
\((\Delta T_{g,j}, \delta t_{g,j})\) are independent for \(i \neq
j\). The fraction of time
contaminated by the first \(n\) glitches is given as
\begin{equation}
f_n = \frac{ \sum_{i=1}^{n} \Delta T_{g,i} }
{\sum_{i=1}^{n} \delta t_{g,i}}.
\end{equation}

Similarly to \cref{sec:normal_approx}, we assume that this fraction
follows a Normal distribution, \(f_n \sim N \left(\mathbb{E}\left[f_n\right],
\mbox{Var}\left[f_n\right]\right)\).
According to \cref{eq:mean_x_y,eq:variance_x_y}, we have
\begin{IEEEeqnarray}{rCl}
\mathbb{E}\left[f_n\right] &\approx& \frac{\mathbb{E}\left[\Delta
T_{g,i}\right]}{\mathbb{E}\left[\delta t_{g,i}\right]},
\\ \mbox{Var}\left[f_n\right] &\approx&
\frac{1}{n} \left(\frac{ \mbox{Var}\left[\Delta T_{g,i}\right] }
{ \mathbb{E}\left[\delta t_{g,i}\right]^2 }  +
\frac{\mathbb{E}\left[\Delta T_{g,i}\right]^2}
{\mathbb{E}\left[\delta t_{g,i}\right]^4} \mbox{Var}\left[\delta
t_{g,i}\right] \right.
\nonumber\\ & &
\left.  - \frac{2\mathbb{E}\left[\Delta
T_{g,i}\right]}{\mathbb{E}\left[\delta
t_{g,i}\right]^3}
\mbox{Cov}\left[\Delta T_{g,i}, \delta t_{g,i}\right]
\right)
\end{IEEEeqnarray}
The PDF of \(\Delta T_{g,i}\) is given by
\begin{IEEEeqnarray}{rCl}
p(\Delta T_{g,i}) &=& \mathrm{e}^{-\lambda \tau} \delta(\Delta T_{g,i} - \tau)
\nonumber\\ && + \lambda \mathrm{e}^{-\lambda \Delta T_{g,i}}
H(\tau-\Delta T_{g,i})H(\Delta T_{g,i}),
\end{IEEEeqnarray}
where \(\delta (x)\) and \(H(x)\) are the Dirac delta and Heaviside
step functions respectively. Using this PDF, we  can evaluate the
expectation and variance of \(\Delta T_{g,i}\)
\begin{align}
\mathbb{E}\left[\Delta T_{g,i}\right]   & = \frac{1}{\lambda} \left( 1
- {\rm e}^{-\lambda \tau} \right), \label{eq:mean_DeltaT} \\
\mbox{Var}\left[ \Delta T_{g,i} \right] & = \frac{1}{\lambda^2} -
\frac{2\tau}{\lambda}
{\rm e}^{-\lambda \tau} - \frac{1}{\lambda^2}{\rm e}^{-2\lambda
\tau}. \label{eq:variance_DeltaT}
\end{align}
The random variable \(\delta t_i\) obeys an exponential distribution with
rate \(\lambda\), so \cref{eq:mean_delta_ti,eq:variance_delta_ti}
hold for it as well.
The joint distribution of \((\Delta T_{g,i}, \delta t_{g,i})\) is given by

\begin{equation}
p(\Delta T_{g,i},\delta t_{g,i}) =
\begin{cases}
\delta(\Delta T_{g,i}-\delta t_{g,i}) \lambda e^{-\lambda \delta
t_{g,i}} \\
\hspace{1.5cm} \text{for } 0 \leqslant \delta t_{g,i} \leqslant \tau, \\
\delta(\Delta T_{g,i}-\tau) \lambda e^{-\lambda \delta t_{g,i}} \\
\hspace{1.5cm} \text{for } \tau \leqslant \delta t_{g,i} < \infty.
\end{cases}
\end{equation}
so the covariance can be computed as
\begin{equation}
\mbox{Cov}\left[\Delta T_{g,i}, \delta t_{g,i}\right]
= \frac{1}{\lambda^2} \left( 1 - \mathrm{e}^{-\lambda \tau} \right)
- \frac{\tau}{\lambda} \mathrm{e}^{-\lambda \tau}.
\end{equation}

Finally, we obtain the Normal approximation to the total contamination
time \(T_\text{ctmn}\) as
\begin{IEEEeqnarray}{rCl}
T_\text{ctmn} &\sim& N \left(\vphantom{T_\text{obs}\left(
1-\mathrm{e}^{-\lambda \tau} \right),
2 \frac{T_\text{obs}}{\lambda} \mathrm{e}^{-\lambda \tau}
\left( 1 - (\lambda \tau + 1)\mathrm{e}^{-\lambda \tau}
\right)}\right. T_\text{obs}\left( 1-\mathrm{e}^{-\lambda \tau} \right),
\nonumber\\ && \hphantom{N\Bigl(\Bigl)} 2
\frac{T_\text{obs}}{\lambda} \mathrm{e}^{-\lambda \tau}
\left( 1 - (\lambda \tau + 1)\mathrm{e}^{-\lambda \tau} \right)
\left.\vphantom{T_\text{obs}\left( 1-\mathrm{e}^{-\lambda \tau} \right),
2 \frac{T_\text{obs}}{\lambda} \mathrm{e}^{-\lambda \tau}
\left( 1 - (\lambda \tau + 1)\mathrm{e}^{-\lambda \tau} \right)}\right),
\end{IEEEeqnarray}
where we replaced \(n\) by its expectation, \(\lambda T_\text{obs}\).
Compared to the general case in \cref{sec:normal_approx}, this
Normal approximation still
remains valid for big \(\tau\). It is the number of artefacts
that needs to be large for
the approximation to hold. However, the above approximation fails
if \(\tau\) is so large that the contamination time
\(T_\text{ctmn}\) approaches the observation time \(T_\text{obs}\),
but we hope that this is an implausible scenario.

\subsection{Merged contamination interval
scenario}\label{sec:merged_ctmn_interval_distribution}
We now return to the Normal approximation to the fraction of the
contamination time as described in \cref{sec:normal_approx} and
consider the first scenario, in which we merge the contamination
intervals if two artefacts happen close to each other.
We need to compute the expectation and variance of \(\Delta T_i\).
A practical way to do this is to consider the PDF \(p(k, T, \tau)\)
where \(k\) is the number of artefacts that fall within a
contamination interval, \(T\) is the total length of the interval and
\(\tau\) is the time separation between the last glitch contributing
to the contamination interval and the end of the
interval, so that \(\tau \leqslant T\).
Let us denote by \(q(t)\) the distribution of the dead time, $t$, of
a random artefact drawn from the population and by $\lambda$ the
Poisson rate of the artefact process. The PDF for \(k=1\) can then be written
\begin{equation}
p(1, T, \tau) = \delta(T - \tau) q(\tau) \mathrm{e}^{-\lambda T}.
\label{eq:pone_mergedint}
\end{equation}
We can determine a recurrence relation for the probability by considering
if the next artefact falls into the current contamination interval:
\begin{IEEEeqnarray}{rCl}
p(k+1, T,&& \tau) = \int_{\tau}^{T} p(k, T, \tau_k) \lambda
\int_0^{\tau} q(\tau^\prime)
\mathrm{d}\tau^\prime \mathrm{d}\tau_k
\nonumber\\ && + \int_0^{T-\tau} \int_{t}^{t+\tau} q(\tau)
\lambda e^{-\lambda (t+\tau-\tau_k)}
\nonumber\\ && \hphantom{\int} \times p(k, T+\tau_k-t-\tau,
\tau_k) \mathrm{d}\tau_k \mathrm{d}t.
\label{eq:recurrence_merged_scenario}
\end{IEEEeqnarray}

\begin{figure}
\includegraphics[width=\linewidth]{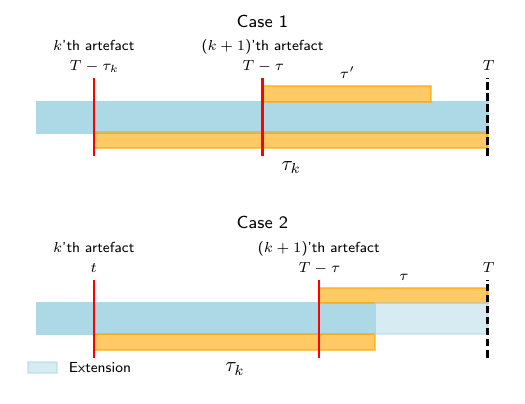}
\caption{
Illustration of the scenarios described by the two different terms in
\cref{eq:recurrence_merged_scenario}.
}
\label{fig:merged_cases}
\end{figure}
The first term of
\cref{eq:recurrence_merged_scenario} corresponds to the case shown
in the upper panel of \cref{fig:merged_cases}, where the
overlapping dead times do not change the contamination interval
(one dead time is inside another). In this case, the $(k+1)$'th
artefact occurs at time $T-\tau$, and contributes dead time $\tau' <
\tau$, while the $k$'th artefact occurs at time $T-\tau_k$ and the
merged dead time contributed by the first $k$ artefacts is $T$. The
second term represents the case shown in the lower panel of
\cref{fig:merged_cases}, in which the $(k+1)$'th artefact,
occurring at time $T-\tau$, increases the length of the interval,
since the amount of dead time contributed by the first $k$ artefacts,
the last of which occurs at time $t$, is $T-\tau<T+\tau_k-t-\tau < T$.
We can then marginalize over \(\tau\) to obtain the PDF for \((k, T)\),
\begin{equation}
p(k, T) = \int_0^T p(k, T, \tau) \, \mathrm{d}\tau,
\label{eq:generic_pdf_k_T}
\end{equation}
and sum \(p(k, T)\) over \(k\) to obtain \(p(T)\), the
PDF for the merged contamination interval \(T\). The expectation
and variance of \(\Delta T_i\)
required for the Normal approximation
in \cref{sec:normal_approx} can then be computed from \(p(T)\).
In principle, we should have an infinite sum over $k$, but in practice,
we do not expect a large number of artefacts to fall close to each
other without corrupting the entire observation time, where the
whole approach breaks down. We truncate the sum at a certain
number (to be determined empirically) depending on the artefact
rate \(\lambda\) and the
distribution of the dead time, \(q(t)\), for the assumed
population of artefacts.

\subsection{Reset contamination interval
scenario}\label{sec:reset_ctmn_interval_distribution}
Here we consider the second scenario, where we reset the
contamination interval if two artefacts fall next to each other.
Following the logic of
\cref{sec:merged_ctmn_interval_distribution}, we see that
\cref{eq:pone_mergedint} remains unchanged, but the recurrence relation
\cref{eq:recurrence_merged_scenario} becomes
\begin{IEEEeqnarray}{rCl}
p(k+1, T,&& \tau) = \lambda q(\tau) \int_{T-\tau}^\infty
\int_{T'-T+\tau}^{T'} p(k,T',\tau')\; \mathrm{d}\tau'
\mathrm{d}T'. \nonumber \\
\label{eq:generic_recurrence_combreset}
\end{IEEEeqnarray}
The probability that a combined contamination interval
has a duration \(T \rightarrow T+\mathrm{d}T\) and
includes \(k\) glitches, \(p(k,T) \mathrm{d} T\), is given by
marginalisation of \cref{eq:generic_recurrence_combreset} over \(\tau\)
as in \cref{eq:generic_pdf_k_T}.


\subsection{Self-contamination}\label{sec:self-contamination}
As mentioned above, here we consider transient GW signals from the
same population which
occur in the data close to each other (``contamination'').  Such
overlapped signals do not necessarily cause loss of
information, therefore, not a real ``contamination'',
but would require more careful (maybe joint) consideration in the
data analysis
\cite{cornish_black_2020,deng_modular_2024,katz_efficient_2024,alvey_what_2023}.
Based on astrophysical predictions for the expected  rate of
observable events, we want to assess the probability of
self-contamination and advise on the necessity of simultaneous
treatment of multiple GW signals in LISA data analysis.


\subsubsection{Constant contamination
period}\label{sec:self-contamination_constant}

The constant dead time $\tau$ associated with each transient GW makes
an analytical treatment of the problem possible. We assume
the case where the contamination works only in one direction with
respect to the signal time of arrival, and if the signal ``A''
affects a signal ``B'', then it affects all the signals occurring between
events ``A'' and ``B''.
The number of contaminated signals is equal to the number of signal
pairs separated by less than
\(\tau\) in time.
This problem reduces to the stick breaking problem given in
\cref{app:stick_breaking}.  We think of the observation time as a
stick of unit length, which is broken into \(n+1\) pieces with
uniformly distributed random break points, where \(n\) is the number of signals.
Following \cref{eq:stick_breaking_exactly_r}, the probability that
exactly \(k\) signals among
\(n\) observed signals are lost to other signals is given by
\begin{IEEEeqnarray}{rCl}
P_{n,k}(T_\text{obs}) &=& \binom{n+1}{k} \left( 1 - (n+1-k)
\frac{\tau}{T_\text{obs}} \right)^{n+1-k}_{+}
\nonumber\\ && \times \sum_{j=0}^{k} (-1)^j \binom{k}{j} \left( 1
- j\frac{\tau}{T_\text{obs}} \right)^{k-1}_{+},
\label{eq:self_contamination_constant_sol}
\end{IEEEeqnarray}
where we define \( x^n_+ = x^n \) if \(x > 0\), \(x^n_+=1\) if \(x
= 0\), and \(x^n_+=0\) if \(x < 0\).
We can connect \cref{eq:self_contamination_constant_sol} to
\(P(k)\), the probability that \(k\) signals are contaminated
during the observation window \(T_\text{obs}\), by summing over
\(n\) which follows a Poisson distribution with
parameter \(\mu T_\text{obs}\).


In practice, it is challenging to accurately numerically compute
\cref{eq:self_contamination_constant_sol}, as binomial coefficients
can be very large. Since the
result \(P_{n,k}(T_\text{obs})\) must lie between
0 and 1, we essentially need to subtract large numbers
to obtain a small number. This can easily lead to numerical errors making
the result unreliable even for moderate values of \(n\) and \(k\).


\subsubsection{Non-constant dead time}
In the general case, where the self-contamination dead time is a random
variable drawn from the uniform or exponential probability
distribution, we can make use of earlier results -- particularly
\cref{eq:generic_pdf_k_T} --
assuming that the occurrence of GW signals also follows a Poisson distribution.
We retain the same definitions of the random variables \(\Delta T_i\)
and \(\delta t_i\) as in \cref{sec:normal_approx}, and denote by \(k_i\)
the number of occurrences of GW signals in the \(i\)'th
contamination interval.
If we assume that only the succeeding signal can be contaminated by
the preceding one, and not vice versa, then the number of
contaminated signals in the \(i\)'th interval is \(k_i - 1\).
If instead we assume that all signals within an interval can
contaminate each other, then the number is
\(f_i k_i\) where \(f_i = 0\) if \(k_i=1\)
and \(f_i = 1\) otherwise. In this section, we consider only the
first case; the second case can be treated similarly and we
indicate below how to obtain it. The random variables $\delta t_i$,
$\Delta T_i$ and $k_i$ are independent for different
\(i\) and have a joint distribution
\begin{equation}
p(\delta t, \Delta T,k)=\mu {\rm e}^{-\mu \delta t} p(k,\Delta T),
\end{equation}
where \(p(k,\Delta T)\) is defined in \cref{eq:generic_pdf_k_T},
after replacing \(\lambda\) by \(\mu\) (the rate of GW signals) in
\cref{eq:recurrence_merged_scenario}.
The occurrence rate of self-contaminated GW signals is given by
\begin{equation}
f_{\rm self}(\{\delta t_i, \Delta T_i, k_i\} : i=1 \ldots n) =
\frac{\sum_{i=1}^n (k_i - 1)}{\sum_{i=1}^n (\delta t_i + \Delta T_i)}.
\label{eq:self_contamination_fraction}
\end{equation}
Defining \(X_i = k_i - 1\) and \(Y_i = \delta t_i + \Delta T_i\),
we can now write down the necessary terms for building the Normal
approximation,
\begin{IEEEeqnarray}{rCl}
\mathbb{E}(X_i) &=& \sum_{k_i=1}^\infty \int_0^\infty (k_i-1)
p(k_i,T)\,{\rm d}T,      \nonumber \\
\mbox{Var}(X_i) &=& \sum_{k_i=1}^\infty \int_0^\infty \left(k_i-1 -
\mathbb{E}(X_i)\right)^2 p(k_i,T)\,{\rm d}T,      \nonumber \\
\mathbb{E}(Y_i) &=& \mathbb{E}(\Delta T_i) + \frac{1}{\mu}, \nonumber \\
\mbox{Var}(Y_i) &=& \frac{1}{\mu^2} + \mbox{Var}(\Delta T_i), \nonumber \\
\mbox{Cov}(X_i, Y_i) &=&  \sum_{k_i=1}^\infty
\int_0^\infty\left(k_i-1 - \mathbb{E}(X_i)\right) \nonumber \\
&& \hphantom{\sum_{k_i=1}^\infty \int} \times  \left(T -
\mathbb{E}(\Delta T_i)\right) p(k_i,T)\,{\rm d}T.
\label{eq:self_contamination_necessary_terms}
\end{IEEEeqnarray}
For the mutual contamination case mentioned above, we need to
replace all the occurrences of \(k_i-1\) by \(f_i k_i\),  and
follow the same reasoning as in the one-sided contamination case,
which we discuss in detail below.

In practice, we cannot always compute the infinite sum over \(k\)
and we need to truncate it at a certain point.  It helps to
introduce additional notation:
\begin{IEEEeqnarray}{rCl}
p_{k_i} &=& \int_0^\infty p(k_i,T)\,{\rm d}T,
\nonumber\\ m_{k_i} &=& \int_0^\infty (k_i-1) p(k_i,T)\,{\rm d}T,
\nonumber\\ \sigma^2_{k_i} &=& \int_0^\infty \left(k_i-1 - m_{k_i}\right)^2
p(k_i,T)\,{\rm d}T,
\nonumber\\ M_{k_i} &=& \int_0^\infty T p(k_i,T)\,{\rm d}T,
\nonumber\\ C_{k_i} &=& \int_0^\infty\left(k_i-1 - m_k\right) \left(T -
M_{k_i}\right) p(k_i,T)\,{\rm d}T,
\end{IEEEeqnarray}
and rewrite the mean and covariance as
\begin{IEEEeqnarray}{rCl}
\mathbb{E}(X_i) &=& m,
\nonumber \\ \mbox{Var}(X_i) &=& \sum_{k_i=1}^\infty \left[
\sigma^2_{k_i} + 2 m^2_{k_i} - m^2_{k_i} p_{k_i} \right] - m^2,
\nonumber \\
\mbox{Cov}(X_i, Y_i) &=&  \sum_{k_i=1}^\infty \left[ C_{k_i} + 2 m_{k_i}
M_{k_i} - m_{k_i} M_{k_i} p_{k_i} \right] - mM,
\nonumber \\[-1em]
\end{IEEEeqnarray}
where \(m = \sum_{k=1}^\infty m_k\) and \(M = \sum_{k=1}^\infty M_k\).
The terms \(m^2\) and \(mM\) can be expanded as Cauchy products,
making it clear that the results are sums of contributions up to
each degree in
$k$. Increasing the degree in $k$ only adds higher-order terms to
the final result,
without affecting the lower-order terms. This implies that the
contributions from higher degrees
in $k$ converge to zero, allowing us to truncate the series to
approximate the sums.
Finally, we 
insert these results into
\cref{eq:mean_x_y,eq:variance_x_y} to obtain
the Normal approximation.

\section{\label{sec:results}Results}
In this section we will numerically compute the distribution of
contamination intervals under the various models described above, and
assess the validity of the Normal approximations. In
\cref{sec:res_ctmn_constant}, we compute the Normal approximation for
the distribution of the total contamination time
\(\pi(T_\text{ctmn})\) introduced in
\cref{eq:unconditional_ctmn_num_pmf} for the case of a
constant dead time and compare it with the simulation results. In
\cref{sec:res_ctmn_merged,sec:res_ctmn_reset}, we show the
analytical results for the PDF \(p(k, T, \tau)\) given in
\cref{eq:recurrence_merged_scenario,eq:generic_recurrence_combreset},
and we compare both the analytical PDF and Normal approximation
for \(\pi(T_\text{ctmn})\) with the numerical simulation.

In
\cref{sec:res_ctmn_constant,sec:res_ctmn_merged,sec:res_ctmn_reset}
we consider contamination caused by glitches similar to
impulse-carrying glitches observed in LISA
Pathfinder data \cite{lisa_pathfinder_collaboration_transient_2022}.
The occurrence of these glitches is well described by a Poisson
process with a rate \SI{1}{\per\day}, which could be tripled, to
\SI{3}{\per\day}, for LISA, given that we will have three identical
spacecrafts.
The classification of three families of impulse-carrying glitches
was proposed in \cite{lisa_pathfinder_collaboration_transient_2022}:
long positive impulse glitches (duration \(\gtrsim
\SI{1.6}{\kilo\second}\)),
short positive impulse glitches (duration \(\lesssim
\SI{1.6}{\kilo\second}\)),
and short negative impulse glitches
(duration \(\lesssim \SI{1.6}{\kilo\second}\)).
The authors of \cite{lisa_pathfinder_collaboration_transient_2022}
identified 30, 51 and 17 glitches within a typical data-taking
window of LISA Pathfinder.
The mean and median durations of long positive impulse glitches are
approximately \SI{9.4}{\kilo\second} and \SI{5.8}{\kilo\second},
respectively. The ratio is close to the ratio, $1/\ln(2)$,
expected for an exponential distribution. This motivates modelling
the glitch durations as
an exponential distribution with a lower cutoff at \(\sim
\SI{1.6}{\kilo\second}\) and mean of \SI{9.4}{\kilo\second}. We will use
this distribution (without the cut-off) in our later analysis.
The durations of short impulse glitches, whether positive or negative,
are usually below \SI{1}{\minute} with only a few outliers near the
cutoff (\(\sim \SI{1.6}{\kilo\second}\)).

The exponential and uniform distributions for the duration of the
glitches used in this paper do not necessarily describe reality very
well; however, they can demonstrate how the final results change
under these different assumptions.
As a constant dead time in \cref{sec:res_ctmn_constant}, we will use
\(\tau=\SI{36}{\second}\) (approximately the median duration of
the short glitches). In
\cref{sec:res_ctmn_merged,sec:res_ctmn_reset}, we will
use an exponential distribution with \(\tau_\text{mean}=\SI{9400}{\second}\)
and a uniform distribution with
\(\tau_\text{max}=\SI{72}{\second}\), with corresponding glitch rates of
\SI{0.92}{\per\day} for the exponential population
and \SI{2.08}{\per\day} for the constant and uniform populations.
These values are consistent with the fraction of long and short
impulse glitches, respectively.


In \cref{sec:res_self_contamination} we consider the
self-contamination of GW signals from MBHBs merging close to each
other in time. The
``contamination'' should be understood to mean that a weak signal
may be undetectable beneath a stronger one, or that significant
biases may arise in parameter estimation when using a single-source
model. We choose the self-contamination dead time to follow an
exponential distribution with a mean of
\(\tau_\text{mean}=\SI{20}{\hour}\), and overlapping dead times are
combined into  common intervals according to the merged scenario.
In reality, the contamination interval depends on the SNR of the
signals, the accumulation of the SNR over time, and the level of
correlation (i.e., overlap in the sense of matched filtering)
between two nearby GW signals.  Depending on the formation model,
we expect anywhere from a few to around 100 MBHB signals per year
in LISA \cite{colpi_lisa_2024}. To illustrate the method, we
arbitrarily choose the rates of \(\mu=\SI{0.05}{\per\day}\) and
\(\mu=\SI{0.2}{\per\day}\), which lie within the expected range and
roughly correspond to heavy and light seed MBH formation channels
\cite{seoane_astrophysics_2023}.



\subsection{Constant dead time scenario}\label{sec:res_ctmn_constant}

\begin{figure}
\includegraphics[width=\linewidth]{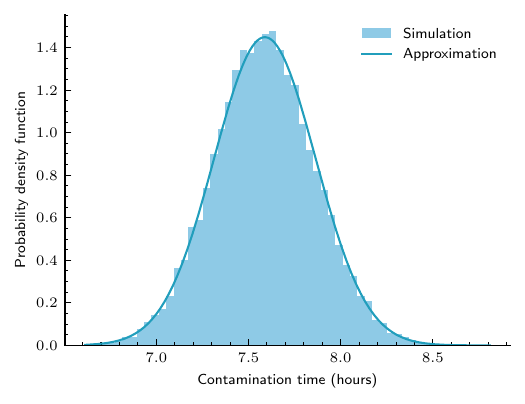}
\caption{Comparison of the PDF for the total contamination time
\(T_\text{ctmn}\) between the Normal approximation
and the simulation in the constant dead time
scenario. \(T_\text{obs} = \SI{365}{\day}\),
\(\lambda = \SI{2.08}{\per\day}\), \(\tau = \SI{36}{\second}\).}
\label{fig:constant_contamination_period}
\end{figure}

Following \cref{sec:fixed_lost_period}, we confirm the validity of
the Normal approximation using numerical simulations. The simulated
data are constructed assuming the artefact rate
\(\lambda=\SI{2.08}{\per\day}\) with a constant dead time
\(\tau=\SI{36}{\second}\), and observation duration
\(T_\text{obs}=\SI{365}{\day}\).  We show
that the Normal approximation (solid line) agrees well with the
numerically evaluated PDF of the total contamination time
\(T_\text{ctmn}\) (histogram) in \cref{fig:constant_contamination_period}

Note that in this case we can also analytically solve for \(P_{n,
l}(T_\text{obs})\), the probability that \(l\) signals are
contaminated by \(n\) artefacts during an observation time \(T_\text{obs}\),
\begin{IEEEeqnarray}{rCl}
P_{n, l}(t) &=& \mathrm{e}^{-\lambda t} \sum_{m=0}^{n}
\sum_{j=0}^{n+l} a^{n, l}_{m, j}
\left( t - m \tau \right)^{j}_{+}
\nonumber\\ && + \mathrm{e}^{-(\mu + \lambda) t} \sum_{m=0}^{n}
\sum_{j=0}^{n+l} A^{n, l}_{m, j}
\left( t - m \tau \right)^{j}_{+},
\label{eq:const_ctmn_analytical_sol}
\end{IEEEeqnarray}
where \(a^{n, l}_{m, j}\) and \(A^{n, l}_{m, j}\) are coefficients
that can be computed
recursively, as shown in \cref{app:constant_ctmn_period}.  It is
straightforward to implement the recursive computation in a
computer program and to obtain the left-hand side of
\cref{eq:unconditional_ctmn_num_pmf} by summing over \(n\),
which itself follows a Poisson distribution with parameter \(\lambda
T_\text{obs}\). However, this is overkill in most practical cases,
as the Normal approximation is already sufficiently accurate.

\subsection{Merged contamination interval scenario}
\label{sec:res_ctmn_merged}
As we discussed in
\cref{sec:normal_approx,sec:merged_ctmn_interval_distribution},
we need the PDF \(p(k, T, \tau)\) to evaluate the mean and covariance
for the Normal approximation.

\subsubsection{Uniform population}\label{sec:res_merged_uniform}
\begin{figure*}[tbh]
\centering
\includegraphics[width=0.48\linewidth]{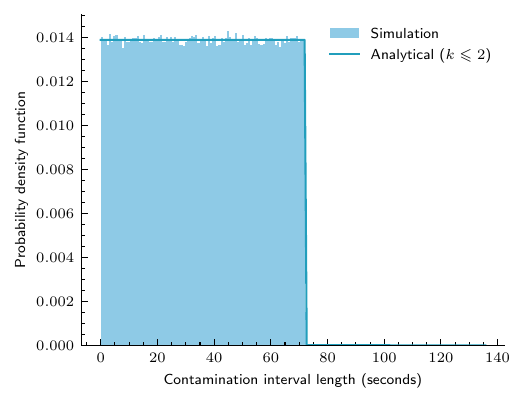}
\includegraphics[width=0.48\linewidth]{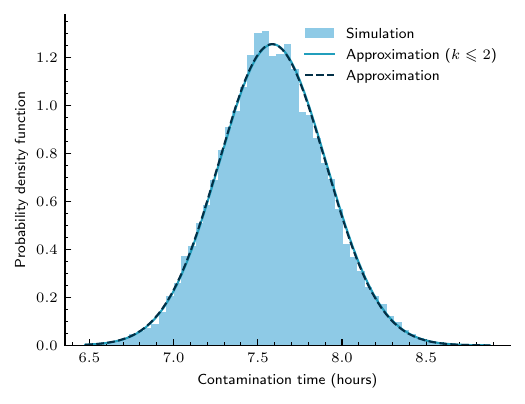}
\caption{Comparison of the PDF for the length of contamination
intervals \(\Delta T_i\) (left)
and the total contamination time \(T_\text{ctmn}\) (right)
between the analytical solution
and the simulation for the uniform contamination population
with the merged interval scenario.
\(T_\text{obs} = \SI{365}{\day}\),
\(\lambda = \SI{2.08}{\per\day}\),
\(\tau_{\text{max}} = \SI{72}{\second}\). The analytical
solution is truncated at \(k=2\).
In the right panel, the Normal approximation computed from the
truncated analytical solution
is shown in blue, while the Normal approximation computed with
the empirical mean and variance
of \(\Delta T_i\) from the simulation is shown in dashed black.}
\label{fig:uniform_contamination_merged_interval}
\end{figure*}

For the dead time distributed uniformly
\(\mathcal{U}[0, \tau_{\text{max}}]\),
we have
\begin{equation}
q(t) = \frac{1}{\tau_{\text{max}}} H(\tau_{\text{max}}-t) H(t).
\end{equation}
We start with $k=1$:
\begin{equation}
p(1, T, \tau) = \delta(T - \tau) \mathrm{e}^{-\lambda T}
\frac{1}{\tau_{\text{max}}} H(\tau_{\text{max}}-\tau),
\label{eq:uniform_ctmn_p1Ttau}
\end{equation}
where we have dropped the term $H(\tau)$ since $\tau$ is always positive.
The marginalized PDF is then given by
\begin{equation}
p(1, T) = \int_0^{T} p(1, T, \tau) \, \mathrm{d}\tau =
\frac{1}{\tau_{\text{max}}} \mathrm{e}^{-\lambda T} H(\tau_{\text{max}}-T).
\end{equation}
Note that for any \(k\), we know that \(p(k, T, \tau) = 0\) if
\(\tau > \tau_\text{max}\).
In the following, we will explicitly add the Heaviside step
function to emphasize the requirement that
\(\tau \leqslant \tau_{\text{max}}\)
even where it is redundant.
Let us now compute \(p(2, T, \tau)\) using the recurrence in
\cref{eq:recurrence_merged_scenario}.
It is helpful to consider different
ranges of \(T\). If \(0 < T \leqslant \tau_{\text{max}}\), then we have
\begin{equation}
p(2, T, \tau) = \frac{2}{\tau_{\text{max}}^2} \lambda
\mathrm{e}^{-\lambda T} \tau H(\tau_{\text{max}}-\tau),
\label{eq:uniform_ctmn_p2Ttau}
\end{equation}
where we used the relation \(\tau \leqslant T\). If \(T >
\tau_{\text{max}}\), then we have
\begin{equation}
p(2, T, \tau) = \frac{1}{\tau_{\text{max}}^2}
\lambda \mathrm{e}^{-\lambda T} \left( \tau - T +
\tau_{\text{max}} \right)^1_+ H(\tau_{\text{max}}-\tau).
\end{equation}
It follows that
\begin{IEEEeqnarray}{rCl}
p(2, T) &=& \int_0^{T} p(2, T, \tau) \, \mathrm{d}\tau
\\ &=& \lambda \mathrm{e}^{-\lambda T}
\frac{T^2}{\tau_{\text{max}}^2} H(\tau_{\text{max}}-T)
\nonumber \\ && +\: \lambda \mathrm{e}^{-\lambda T}
\frac{1}{2\tau_{\text{max}}^2} (T-2\tau_{\text{max}})^2
\nonumber \\ && \quad \times \, H(-\tau_{\text{max}}+T)
H(2\tau_{\text{max}}-T).
\end{IEEEeqnarray}
It becomes challenging to compute \(p(3, T)\) in this
way due to a large number of terms involved.
Moreover, it is possible to show that for  \(k \geqslant 2\), the
PDF takes the form
\begin{IEEEeqnarray}{rCl}
p&&(k,T,\tau) = \lambda^{k - 1} \tau_\text{max}^{- k} e^{- T
\lambda} H\left(- \tau + \tau_\text{max}\right)
\sum_{\substack{0 \leqslant l \leqslant j\\0 \leqslant j \leqslant 2k-3}}
\nonumber \\ &&
\left[\vphantom{\sum_{\substack{0 \leqslant q \leqslant l\\1
\leqslant m \leqslant k}} T^{j - l} \tau^{- j + 2 k - 3}
\tau_\text{max}^{l - q} (- T + m \tau_\text{max})^{q}_{\!+}
{D}^{l,q,m}_{j,k}
+ \sum_{\substack{0 \leqslant q \leqslant l\\0 \leqslant m
\leqslant k}} T^{j - l} \tau^{- j + 2 k - 3}
\tau_\text{max}^{l - q} (- T + m \tau_\text{max} +
\tau)^{q}_{\!+} {C}^{l,q,m}_{j,k}}\right.
\sum_{\substack{0 \leqslant q \leqslant l\\0 \leqslant m
\leqslant k}} T^{j - l} \tau^{- j + 2 k - 3}
\tau_\text{max}^{l - q} (- T + m \tau_\text{max} +
\tau)^{q}_{\!+} {C}^{l,q,m}_{j,k}
\nonumber \\ && \hphantom{.} + \sum_{\substack{0 \leqslant q
\leqslant l\\1 \leqslant m \leqslant k}} T^{j - l} \tau^{- j + 2 k - 3}
\tau_\text{max}^{l - q} (- T + m \tau_\text{max})^{q}_{\!+}
{D}^{l,q,m}_{j,k}
\left.\vphantom{ T^{j - l} \tau^{- j + 2 k - 3}
\tau_\text{max}^{l} {E}_{l,j,k}
+ \sum_{\substack{0 \leqslant q \leqslant l\\1 \leqslant m
\leqslant k}} T^{j - l} \tau^{- j + 2 k - 3}
\tau_\text{max}^{l - q} (- T + m \tau_\text{max})^{q}_{\!+}
{D}^{l,q,m}_{j,k}
+ \sum_{\substack{0 \leqslant q \leqslant l\\0 \leqslant m
\leqslant k}} T^{j - l} \tau^{- j + 2 k - 3}
\tau_\text{max}^{l - q} (- T + m \tau_\text{max} +
\tau)^{q}_{\!+} {C}^{l,q,m}_{j,k}}\right],
\nonumber \\ &&
\label{eq:uniform_ctmn_merged_pdf}
\end{IEEEeqnarray}
where the coefficients \({C}^{l,q,m}_{j,k}\) and
\({D}^{l,q,m}_{j,k}\) can be computed recursively through
\cref{eq:recurrence_merged_scenario} with known initial conditions
determined by \cref{eq:uniform_ctmn_p2Ttau}.
It is still tedious to derive the recurrence relations for these
coefficients, and the numerical implementation is prone to errors,
similar to what we describe in
\cref{sec:self-contamination_constant,app:analytical_solution}.
We do not present the analytical solution in this work, as the
number of terms is very large.
In \cref{fig:uniform_contamination_merged_interval}, we compare the
PDF of the contamination intervals \(\Delta T_i\) and the total
contamination time \(T_\text{ctmn}\)
between the analytical solution and the simulation for the uniform
contamination population with the scenario of the merged interval.
The analytical solution
is truncated at (k=2), and we observe that the blue curve in the
right panel still agrees well with the simulation. The dashed black
curve in the right panel shows the normal approximation computed
from the empirical mean and variance of \(\Delta T_i\)
based on the simulation data in the left panel.
This serves as a reference for assessing the appropriateness of
the truncated analytical solution, as the difference is determined by
the terms with $k > 2$.

The reason for the good agreement despite the relatively low
truncation order is that, due to the short maximal dead time
(\SI{72}{\second}) and a moderate glitch rate
(\SI{2.08}{\per\day}), the probability of having more than two
signals in a single contamination interval is small.
As a result, we expect the contribution from \(k \geqslant 3\) to be
negligible, and this is confirmed by the simulation results.
The left panel of \cref{fig:uniform_contamination_merged_interval}
shows the PDF of the length of the contamination intervals \(\Delta
T_i\), which appears to be a uniform distribution in \([0,
\tau_{\text{max}}]\).
This is again due to our choice of parameters, which makes the
probability of merging contamination intervals so small that any
deviation from the uniform distribution is hardly noticeable:
we have
\(
p(1) \approx 0.99913\) and \(p(2) \approx 8.65\times10^{-4}
\)
where \(p(k):=\int_0^{\infty} p(k,T)\,\mathrm{d}T\).

\subsubsection{Exponential population}\label{sec:res_merged_exponential}
Now let us consider an exponential dead time
distribution \(\mathcal{E}(\tau_{\text{mean}})\).
We follow similar steps as in \cref{sec:res_merged_uniform} to
compute the PDF \(p(k, T, \tau)\) and
then integrate over \(\tau\).
We have
\begin{equation}
q(t) = \nu {\rm e}^{-\nu t}, \label{eq:exponential_q}
\end{equation}
where \(\nu := 1/\tau_\text{mean}\) is the rate of the exponential
distribution.
The first three terms are obtained from the recurrence relation
in \cref{eq:recurrence_merged_scenario}, followed by integration
over \(\tau\):
\begin{align}
p(1,T) & =\nu {\rm e}^{-(\lambda+\nu)T} \nonumber
\\
p(2,T) & = 2\lambda \nu {\rm e}^{-(\lambda+\nu)T} \left(T -
\frac{1}{\nu}(1-{\rm e}^{-\nu T})\right) \nonumber
\\
p(3,T) & = 2\lambda^2 \nu {\rm e}^{-(\lambda+\nu)T} \left[ T^2 +
\frac{T}{\nu} \left({\rm e}^{-\nu T} - \frac{5}{2}\right)
\right. \nonumber \\
& \hspace{1cm} \left.- \frac{3}{\nu^2} {\rm e}^{-\nu T} +
\frac{3}{4\nu^2}{\rm e}^{-2\nu T} + \frac{9}{4\nu^2}\right].
\end{align}

It is somewhat more manageable to compute higher-order terms in
\(k\) compared to
\cref{sec:res_merged_uniform},
and we can also derive a general expression for  \(p(k, T, \tau)\)
for \(k \geqslant 2\) of the form
\begin{IEEEeqnarray}{rCl}
p(&&k,T,\tau) = \lambda^{k - 1} \tau_\text{mean}^{k - 3} e^{- T \lambda}
\nonumber \\ && \times \sum_{\substack{0 \leqslant l \leqslant j
\\0 \leqslant j \leqslant k\\1 \leqslant m \leqslant k
\\1 \leqslant q \leqslant k}}
T^{l} \tau^{j - l} \tau_\text{mean}^{- j} e^{- \frac{T m}{\tau_\text{mean}}}
e^{- \frac{\tau (q-m)}{\tau_\text{mean}}} {F}_{j,l,m,q},
\label{eq:exponential_ctmn_analytical_sol}
\end{IEEEeqnarray}
where the coefficients \({F}_{j,l,m,q}\) themselves satisfy a
recurrence relation.
We present the analytical result in
\cref{app:exponential_ctmn_period}, but this solution is hardly
practical due to numerical inaccuracy in the implementation, as
discussed in \cref{app:exponential_ctmn_period}.

\begin{figure*}[tbh]
\centering
\includegraphics[width=0.48\linewidth]{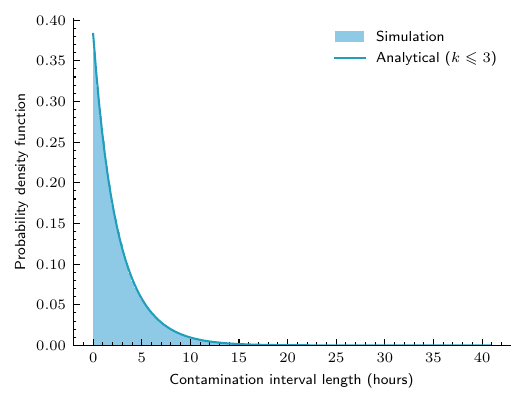}
\includegraphics[width=0.48\linewidth]{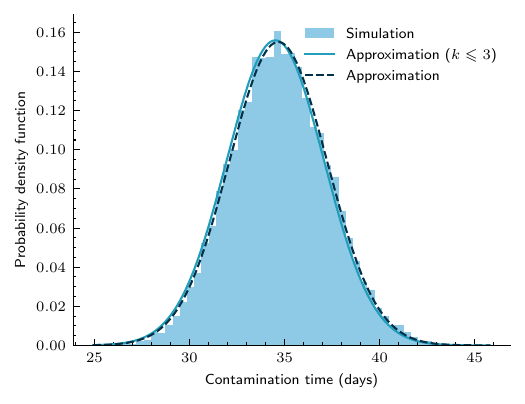}
\caption{Comparison of the PDF for the length of contamination
intervals \(\Delta T_i\) (left)
and the total contamination time \(T_\text{ctmn}\) (right)
between the analytical solution
and the simulation for the exponential contamination population
with the merged interval scenario.
\(T_\text{obs} = \SI{365}{\day}\),
\(\lambda = \SI{0.92}{\per\day}\),
\(\tau_{\text{mean}} = \SI{9400}{\second}\). The analytical
solution is truncated at \(k=3\).
In the right panel, the Normal approximation computed from the
truncated analytical solution
is shown in blue, while the Normal approximation computed with
the empirical mean and variance
of \(\Delta T_i\) from the simulation is shown in dashed black.}
\label{fig:exponential_contamination_merged_interval}
\end{figure*}

In \cref{fig:exponential_contamination_merged_interval}, we compare
the PDF of the contamination intervals \(\Delta T_i\) and the total
contamination time \(T_\text{ctmn}\) between the analytical
solution and the simulation for the exponential contamination
population in the scenario of the merged interval. The analytical
solution is truncated at \(k=3\) and shows a slight deviation from
the empirical normal approximation (which uses the empirically
computed mean and variance) in the right panel of
\cref{fig:exponential_contamination_merged_interval}. This
discrepancy is caused by the truncation at \(k=3\) which is
insufficient to fully capture the distribution of contamination
intervals \(\Delta T_i\) when the mean dead time is long
(\SI{9400}{\second}). Note that this difference is not visible in
the left panel.

\subsection{Reset contamination interval scenario}\label{sec:res_ctmn_reset}

Similarly to \cref{sec:res_ctmn_merged}, we need the PDF  \(p(k, T,
\tau)\) where the recurrence is now given by
\cref{eq:generic_recurrence_combreset},
in order to establish the Normal approximation.

\subsubsection{Uniform population}
In the reset scenario, \(p(1, T, \tau)\) remains the same as in
\cref{eq:uniform_ctmn_p1Ttau}.
We then compute \(p(2, T, \tau)\) and integrate over \(\tau\) to get
\begin{IEEEeqnarray}{rCl}
p(1,T) &=& \frac{{\rm e}^{-\lambda T}}{\tau_\text{max}} H(T)
H(\tau_\text{max} - T), \nonumber \\
p(2,T) &=& \frac{\lambda {\rm e}^{-\lambda T}
T^2}{\tau_\text{max}^2} \left(\frac{\tau_\text{max}}{T}
- \frac{1}{2}\right) H(T) H(\tau_\text{max} - T) \nonumber \\
&& + \frac{\lambda {\rm e}^{-\lambda T} T^2}{\tau_\text{max}^2}
\frac{1}{2} \left[4 \left(\frac{\tau_\text{max}}{T}\right)^2
- 4 \frac{\tau_\text{max}}{T} + 1 \right] \nonumber \\
&& \times H(T - \tau_\text{max}) H(2\tau_\text{max} - T).
\end{IEEEeqnarray}
In this scenario it is manageable to compute \(p(3, T, \tau)\) and
we find it to be
\begin{IEEEeqnarray}{rCl}
p(3,T) &=& \frac{\lambda^2 {\rm e}^{-\lambda T}
T^4}{\tau_\text{max}^3} \frac{1}{24} \left[12
\left(\frac{\tau_\text{max}}{T}\right)^2
- 8 \frac{\tau_\text{max}}{T} + 1 \right] \nonumber \\
&& \times H(T) H(\tau_\text{max} - T) \nonumber \\
&& + \frac{\lambda^2 {\rm e}^{-\lambda T} T^4}{\tau_\text{max}^3}
\frac{1}{24} \left[-31 \left(\frac{\tau_\text{max}}{T}\right)^4
+ 84 \left(\frac{\tau_\text{max}}{T}\right)^3 \right. \nonumber \\
&& \hphantom{+} \left.-66
\left(\frac{\tau_\text{max}}{T}\right)^2 + 20
\frac{\tau_\text{max}}{T} - 2\right] \nonumber \\
&& \times H(T - \tau_\text{max}) H(2\tau_\text{max} - T) \nonumber \\
&& + \frac{\lambda^2 {\rm e}^{-\lambda T} T^4}{\tau_\text{max}^3}
\frac{1}{24} \left[81 \left(\frac{\tau_\text{max}}{T}\right)^4
- 108 \left(\frac{\tau_\text{max}}{T}\right)^3 \right. \nonumber \\
&& \hphantom{+} \left.54 \left(\frac{\tau_\text{max}}{T}\right)^2
- 12 \frac{\tau_\text{max}}{T} + 1\right] \nonumber \\
&& \times H(T - 2\tau_\text{max}) H(3\tau_\text{max} - T).
\end{IEEEeqnarray}

\begin{figure*}[tbh]
\centering
\includegraphics[width=0.48\linewidth]{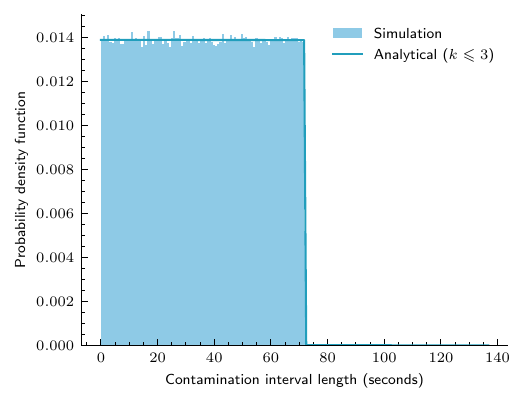}
\includegraphics[width=0.48\linewidth]{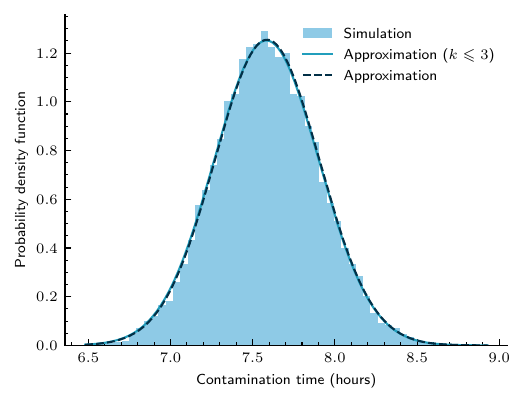}
\caption{Comparison of the PDF for the length of contamination
intervals \(\Delta T_i\) (left)
and the total contamination time \(T_\text{ctmn}\) (right)
between the analytical solution
and the simulation for the uniform contamination population
with the reset interval scenario.
\(T_\text{obs} = \SI{365}{\day}\),
\(\lambda = \SI{2.08}{\per\day}\),
\(\tau_{\text{max}} = \SI{72}{\second}\). The analytical
solution is truncated at \(k=3\).
In the right panel, the Normal approximation computed from the
truncated analytical solution
is shown in blue, while the Normal approximation computed with
the empirical mean and variance
of \(\Delta T_i\) from the simulation is shown in dashed black.}
\label{fig:uniform_contamination_reset_interval}
\end{figure*}

In \cref{fig:uniform_contamination_reset_interval}, we compare the
PDF of the contamination intervals
\(\Delta T_i\) and the total contamination time \(T_\text{ctmn}\)
between the analytical solution and the simulation for the uniform
contamination population in the reset interval scenario. Since we
were able to include the term \(k=3\) in the analytical solution,
it is not surprising that the analytical result matches well the
simulation for the short maximal dead time (\SI{72}{\second}),
which we had already achieved with the truncation \(k=2\) in the
merged scenario shown in \cref{fig:uniform_contamination_merged_interval}.

\subsubsection{Exponential population}
In the reset scenario with the exponential dead time distribution
(\cref{eq:exponential_q}),
we can compute \(p(k, T, \tau)\) for all \(k\), assuming \(T
\geqslant 0\). This is
\begin{IEEEeqnarray}{rCl}
p(1,T,\tau) &=& \delta(T-\tau) \nu {\rm e}^{-(\lambda+\nu)T},
\nonumber    \\
p(k,T,\tau) &=& \frac{\lambda^{k-1} \nu}{(k-2)!} (T-\tau)^{k-2}
{\rm e}^{-(\lambda+\nu)T}\;\text{for}\; k \geqslant 2. \nonumber \\
\end{IEEEeqnarray}
Then, summing over \(k\), we obtain
\begin{align}
p(T,\tau) & = \lambda \nu {\rm e}^{-\lambda \tau} {\rm e}^{-\nu
T} + \nu {\rm e}^{-(\lambda+\nu)T} \delta (T-\tau), \nonumber \\
p(T)      & = \int_0^T p(T,\tau) {\rm d}\tau = \nu {\rm e}^{-\nu T}.
\end{align}
Interestingly, the distribution of the contamination intervals
\(\Delta T_i\) is distributed in exactly the same way as the
exponential distribution of dead times in this case. This property is
peculiar to the exponential distribution of dead times and its
relation to the spacing of events drawn from a Poisson process. Using
an alternative continuous distribution, for example a Gamma
distribution with shape parameter different from one, leads to a
distribution of contamination intervals that differs from the
underlying dead time distribution. 

\begin{figure*}[tbh]
\centering
\includegraphics[width=0.48\linewidth]{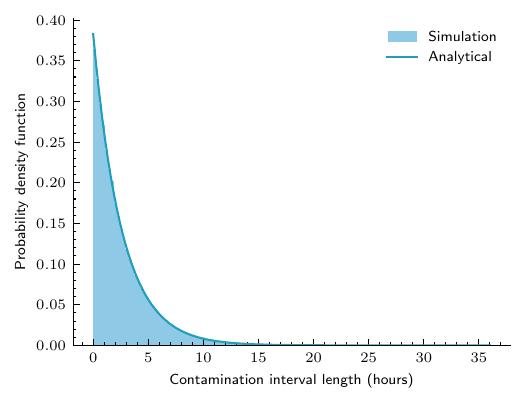}
\includegraphics[width=0.48\linewidth]{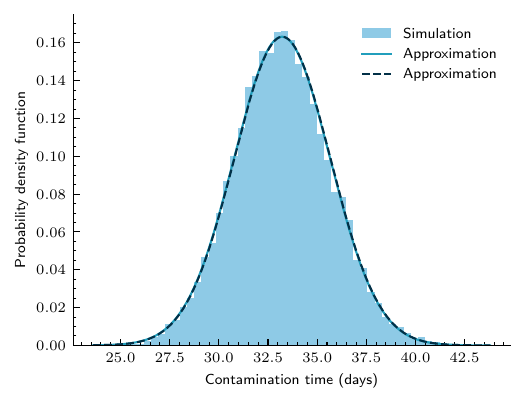}
\caption{Comparison of the PDF for the length of contamination
intervals \(\Delta T_i\) (left)
and the total contamination time \(T_\text{ctmn}\) (right)
between the analytical solution
and the simulation for the exponential contamination population
with the reset interval scenario.
\(T_\text{obs} = \SI{365}{\day}\),
\(\lambda = \SI{0.92}{\per\day}\),
\(\tau_{\text{mean}} = \SI{9400}{\second}\). The analytical
solution is not truncated.
In the right panel, the Normal approximation computed from the
analytical solution
is shown in blue, while the Normal approximation computed with
the empirical mean and variance
of \(\Delta T_i\) from the simulation is shown in dashed black.}
\label{fig:exponential_contamination_reset_interval}
\end{figure*}

In \cref{fig:exponential_contamination_reset_interval} we compare
the PDF of the contamination intervals \(\Delta T_i\) and the total
contamination time \(T_\text{ctmn}\)
between the analytical solution and the simulation for the
exponential contamination population in the reset interval scenario.
The analytical solution is not truncated and we observe a perfect
match with the simulation.



\subsection{Self-contamination}\label{sec:res_self_contamination}



\begin{figure*}[tbh]
\centering
\includegraphics[width=0.48\linewidth]{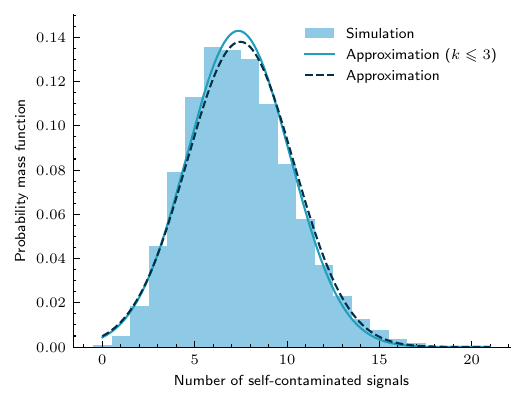}
\includegraphics[width=0.48\linewidth]{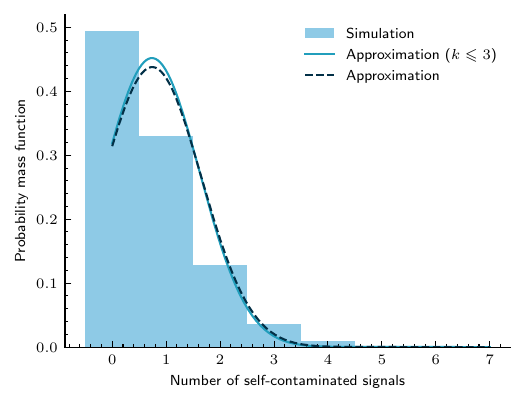}
\includegraphics[width=0.48\linewidth]{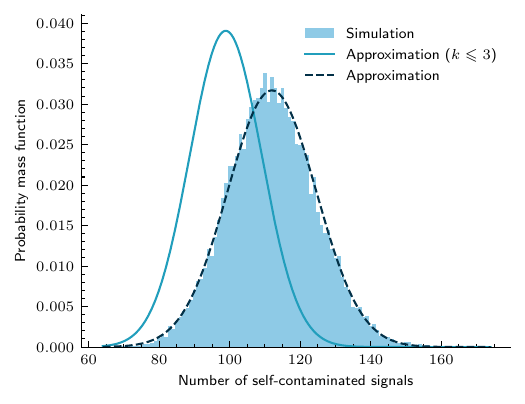}
\includegraphics[width=0.48\linewidth]{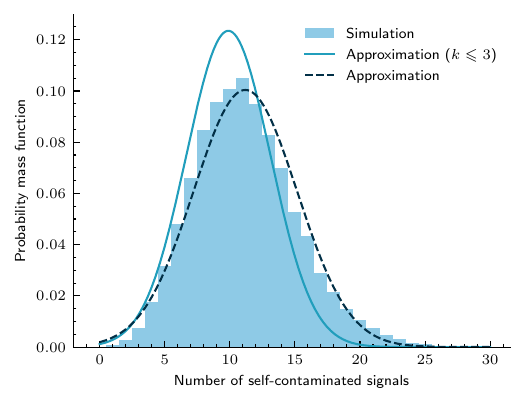}
\caption{Comparison between the Normal approximation and the
simulated distribution for the number of self-contaminated signals.
The self-contamination dead times are drawn from an exponential
distribution of \(\tau_{\text{mean}} = \SI{20}{\hour}\)
and the contamination intervals are obtained in the merged scenario.
\(\mu = \SI{0.05}{\per\day}\) (upper panels)
\(\mu = \SI{0.2}{\per\day}\) (lower panels),
\(T_\text{obs} = \SI{3650}{\day}\) (left panels),
\(T_\text{obs} = \SI{365}{\day}\) (right panels).
The analytical solution is truncated at \(k=3\). The black
dashed curve is the Normal approximation
computed from the empirical mean, variance and covariance (see
\cref{eq:self_contamination_necessary_terms}) of the simulated data.}
\label{fig:self_contamination_exponential}
\end{figure*}

In \cref{fig:self_contamination_exponential}
we compare the Normal approximation and the simulation for the
distribution of the number of self-contaminated signals.
The analytic Normal approximation is truncated at $k\le 3$ and
given by a solid blue line, while the Normal distribution with
empirical mean and variance (estimated from the simulation
according to
\cref{eq:self_contamination_fraction,eq:mean_x_y,eq:variance_x_y})
is given by a dashed black line.

We have considered two astrophysical populations described above
with rate \(\mu = \SI{0.05}{\per\day}\) (upper panels) and \(\mu =
\SI{0.2}{\per\day}\) (lower panels). We also took two observational
periods: 1 year \(T_\text{obs} = \SI{365}{\day}\) (right panels)
and 10 years \(T_\text{obs} = \SI{3650}{\day}\) (left panels).
Solid histograms demonstrate numerical results on the simulated data.

The discrepancy between the Normal approximation and the numerical
distribution can be
partially attributed to small number statistics, which is especially evident in
the upper right plot (low rate and 1 year of observation).
In three plots (besides the lower left) the distributions are
truncated at zero, causing a slight bias in the Normal
approximation. Despite that, the empirical Normal distributions fit
the numerical results rather well in all the scenarios considered.
Finally, the discrepancy with the truncated Normal approximation is
caused by the truncation at $k=3$.
Given our setup, self-contamination of several signals is
rather common and this is clear in the lower left panel in which it
is obvious that the Normal approximation should be valid, but there
is a big difference between the two different Normal approximations.
By contrast, in the upper panels, there is good agreement between the
truncated and empirical Normal approximations, indicating that
intervals with more than three overlapping sources are sufficiently
rare. 

\subsection{Put constraints on the contamination rate and
duration}\label{sec:constraints}
In this section, we apply our results to set requirements on
artefacts that occur during LISA's
operation time.  The inputs characterizing the artefacts are the
contamination rate $\lambda$ and the characteristic dead-time
parameter, with its associated distribution: \(\tau\) for the
constant population, \(\tau_{\text{mean}}\) for the exponential population,
and \(\tau_{\text{max}}\) for the uniform population.
Given a mission duration, we want to set a threshold
(\(T_\text{critical}\)) for the total contamination time. Then, we
adopt an analogue of a survival function  -- more appropriately
called a \emph{death function} (DF) in our context -- defined as
the probability that the total contamination time is greater
than the critical threshold:
$$\text{DF} = P(T_\text{ctmn} \ge T_\text{critical}).$$
We estimate the DF on a two-dimensional grid of artefact rate and
characteristic dead time. This will serve as a figure of merit
which will demonstrate which population of artefacts could
potentially jeopardize LISA's scientific objectives.
We compute this figure of merit using the Normal approximation
described in this paper, and it comes with the restriction of
large-sample statistics, although the approach does also give a
reasonable approximation in the regime of low numbers of events; see
\cref{sec:res_self_contamination}. This implicitly makes some
imposes some constraints on the mission duration \(T_\text{obs}\) and the
parameters of the artefact population. As for the choice of
\(T_\text{critical}\), it should not be close to the
\(T_\text{obs}\), since (i)  we want to avoid excessive
contamination;and (ii) such proximity could violate the assumptions
of our approximation.

\begin{figure*}
\centering
\includegraphics[width=0.48\linewidth]{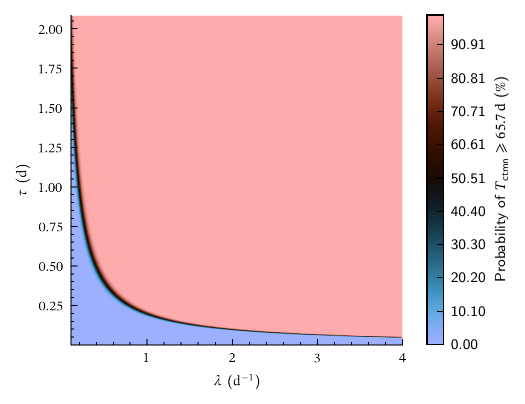}
\includegraphics[width=0.48\linewidth]{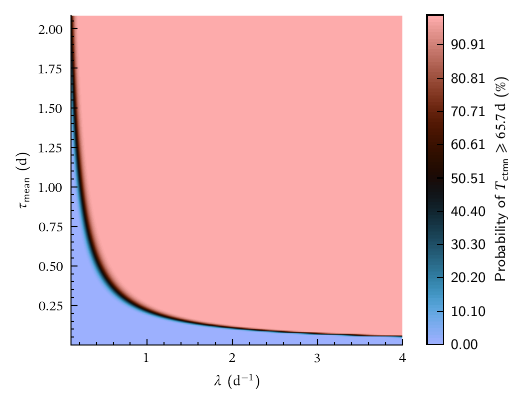}
\includegraphics[width=0.48\linewidth]{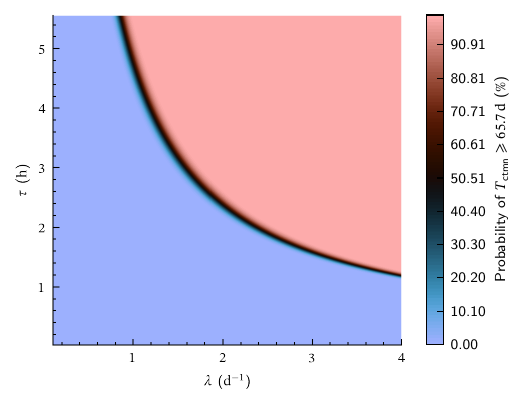}
\includegraphics[width=0.48\linewidth]{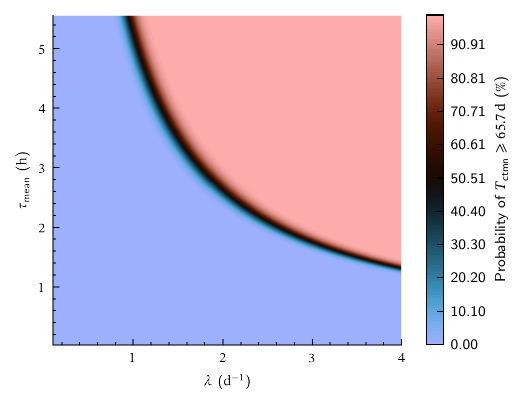}
\caption{Contour region plot of the survival function
\(\text{SF}(T_\text{critical})\) as a function of the contamination rate,
\(\lambda\), and the contamination duration parameter, \(\tau\)
(left panels) or \(\tau_{\text{mean}}\) (right panels).
The lower panels show the same contour plots but for a smaller
range of the rate and a larger range of the duration parameter.
\(T_\text{critical}=\SI{65.7}{\day}\), \(T_\text{obs} = \SI{365}{\day}\).}
\label{fig:contour_plots}
\end{figure*}

We demonstrate the merit figure in \cref{fig:contour_plots} as
contour plots of the DF. The left panel corresponds to the constant
dead-time distribution, while the right panel represents the
exponential distribution in the reset contamination interval
scenario. The lower panels show a zoomed-in view for small dead-time
values. We set \(T_\text{critical}=\SI{65.7}{\day}\) as an
illustrative value for a one-year observation, corresponding to an
82\% duty cycle; our approach, however, allows the DF to be computed
for any \(T_\text{critical}\) and \(T_\text{obs}\). The figures reveal
a sharp transition in probability,
enabling a clear identification of the boundary beyond which artefact
populations become ``dangerous'' to the mission.

\section{\label{sec:summary}Summary}
In this work we demonstrate how to compute the probability that
transient GW signals are contaminated by
various artefacts (gaps and glitches). We also discuss
self-contamination: the possibility of observing
overlapping merging MBHB signals that occur too close to each other in time.
We have derived a Normal approximation to describe the distribution
of the total contamination time \(T_\text{ctmn}\)
and the number of self-contaminated signals in the
self-contamination case.  This approximation was compared to the
results of simulations, and we confirmed its validity for realistic
parameters across several populations of artefacts. The populations
are defined by a distribution of the dead time associated with each
artefact (constant, uniform, exponential). In addition, we considered
two scenarios for building a contamination interval when dead times
overlap: (i) the merged scenario (the interval is constructed as
the union of two dead times); and (ii) the reset scenario (where the
last artefact defines the end time of the contamination period).
For each case, we have demonstrated how to construct the mean and
variance of the Normal approximation
using a recurrence, which we truncated for practical reasons. We
have validated the approximation by comparing to the numerical
simulations based on parameter choices informed by LISA Pathfinder
data,  highlighting both the
limitations of the Normal assumption and discrepancies introduced
by truncating the number of overlapping dead times within a single
contamination interval.  Finally, we proposed a figure of merit to
assess the impact of a given artefact population on the scientific
performance of the LISA mission.

The analysis we have done here is based on a relatively simple model of
glitch populations, which we believe can nonetheless be used to
assess the impact of glitches on the scientific data produced by any
experiment, and in particular LISA. As the LISA instrument is
developed over the coming years, and our understanding of likely
sources of glitches progresses, the formalism developed here, and the
associated code, can be used to quickly assess the impact of new
glitch populations and flag any that are dangerous for the scientific
output of the mission. We hope that this work will provide a useful
figure of merit tool for assessing future mission design choices and
ensure that the LISA mission at launch generates data of high quality
for science.


\begin{acknowledgments}
The authors acknowledge support from the CNES for the exploration of
LISA science.
S.D.\ acknowledges financial support from CNES.
The participation of S.B.\ in this project has received funding from
the European Union’s
Horizon 2020 research and innovation program under the Marie Skłodowska-Curie
grant agreement No.\ 101007855.
S.B.\ acknowledges funding from the French National Research Agency
(grant ANR-21-CE31-0026, project MBH\_waves).
Numerical computations were partly performed on the DANTE platform, APC, France.
The figures were created with LovelyPlots
\cite{sheriff_lovelyplots_2022} and Matplotlib \cite{Hunter:2007}.
\end{acknowledgments}

\bibliography{references}
\appendix
\section{Stick breaking solution}\label{app:stick_breaking}
Let us consider a stick of unit length, which is broken into \(n+1\)
pieces with uniformly distributed random break points. We can prove
by induction that the probability that the
first \(r\) pieces of the broken stick all exceed length
\(x\) is given by
\begin{equation}
p_{n,r} = \left( 1 - rx \right)^{n}_{+}. \label{eq:stick_breaking_first_r}
\end{equation}
In the case \(n=1\), there is one break point and we have
either \(r=1\) or \(r=2\) (the case \(r=0\) is trivial). For
\(r=1\), the probability is just the probability
that the break point is in the interval \(\left[x, 1\right]\) which
evaluates to \((1-x)_+\). For \(r=2\), the
probability is the probability that the break point is in the
interval \(\left[x, 1-x\right]\), which evaluates to
\((1-2x)_{+}\). The result for \(n=1\) then follows. Now for the
induction step let us suppose the result holds
for \(n=k\) and consider the case \(n=k+1\). The probability that
the first break point is in the interval
\(\left[u, u+ \mathrm{d}u \right]\) is \( (k+1) \mathrm{d}u
(1-u)^{k} \) which is the number of ways to choose
the first break point from the set of \(k+1\) break points, times
its probability density, times the probability
that the remaining break points all lie in the interval \(\left[u,
1\right]\). The first \(r\) pieces will all
exceed $x$ if the first break point is at \(u \in \left[x, 1\right]\)
and the remaining \(r-1\) pieces all
exceed \(x\). The latter is the probability that a stick of length
\(1-u\) broken into \(k\) pieces has the first \(r-1\) pieces all
exceeding \(x\), which is \(\left( 1 -
(r-1)x/(1-u) \right)^{k}_{+}\) by the induction
hypothesis. Note that we have scaled both the length of the stick and
the pieces by \(1-u\). It follows that
\begin{equation}
\begin{aligned}
p_{k+1,r} & = \int_{x}^{1} (k+1) (1-u)^{k} \left( 1 -
(r-1)\frac{x}{1-u} \right)^{\!k}_{+} \, \mathrm{d}u
\\ & = \int_{x}^{1-(r-1)x} (k+1) (1-u-(r-1)x)^{k} \,\mathrm{d}u
\\ & = \left( 1 - rx \right)^{k+1}_{+},
\end{aligned}
\end{equation}
so that the result holds for \(n=k+1\) and result
\cref{eq:stick_breaking_first_r} holds by induction.

We can now compute the probability that the maximum length of the
first \(m+1\) pieces
does not exceed \(x\). The statement that the \(r\)'th piece is
shorter than \(x\) is the complement
of the statement that the \(r\)'th piece is longer than \(x\). We
denote the latter by \(X_r\) and the
former by \(\bar{X}_r\). The probability we seek is then
\cref{eq:stick_breaking_max_prob}.
As we are free to reorder the pieces, the probabilities are
independent of the labeling of the pieces, which leads to
\cref{eq:stick_breaking_max_prob_solution}.

Finally, we give the probability that exactly \(r\) pieces exceed length
\(x\). This is given by the probability
that the first \(r\) pieces exceed length \(x\) and the maximum
length of the remaining \(n-r+1\) pieces does not
exceed \(x\), which is \cref{eq:stick_breaking_exactly_r}.

\begin{widetext}
\begin{IEEEeqnarray}{rCl}
\mathbb{P}\left( \bar{X}_1 \cap \bar{X}_2 \cap \ldots \cap
\bar{X}_{m} \cap \bar{X}_{m+1} \right)
&=& 1 - \mathbb{P}(X_1) - \cdots - \mathbb{P}(X_{m+1}) +
\mathbb{P}(X_1 \cap X_2)
\nonumber\\ && + \cdots + \mathbb{P}(X_{m} \cap X_{m+1})
+ \cdots
+ (-1)^{m+1} \mathbb{P}(X_1 \cap \cdots \cap X_{m+1})
\label{eq:stick_breaking_max_prob}
\end{IEEEeqnarray}
\begin{equation}
\mathbb{P}\left( \bar{X}_1 \cap \bar{X}_2 \cap \ldots \cap
\bar{X}_{m} \cap \bar{X}_{m+1} \right)
= \sum_{j=0}^{m+1} (-1)^j \binom{m+1}{j} \left( 1 - jx \right)^{m}_{+}
\label{eq:stick_breaking_max_prob_solution}
\end{equation}
\begin{equation}
\mathbb{P}_{n,r}(x) = \binom{n+1}{r} \left( 1 - rx \right)^{r}_{+}
\sum_{j=0}^{n-r+1} (-1)^j \binom{n-r+1}{j} \left( 1 - jx \right)^{n-r}_{+}
\label{eq:stick_breaking_exactly_r}
\end{equation}
\end{widetext}

\hphantom{} \clearpage
\section{Analytical solution to the integral recurrence
relations}\label{app:analytical_solution}

\subsection{Constant dead time}\label{app:constant_ctmn_period}
As mentioned in \cref{sec:res_ctmn_constant}, in this case it is
possible to analytically compute the
probability that exactly \(n\) glitches (happening at rate \(\lambda\)) occur
and \(l\) signals (happening at rate \(\mu\))
are lost to them in an observation window of length \(T\), which we
denote by \(P_{n,l}(T)\).
We have trivially \(P_{0,l}(T) = \mathrm{e}^{-\lambda T}\) for all
\(l\) and \(T \geqslant 0\).

With the definition \(P_{n,l}(T)=0\) for \(T < 0\), we can derive
the recurrence relation \cref{eq:iterative_two_populations} between
the indexed probabilities.
It is easy to show that the general solution to
\cref{eq:iterative_two_populations} is given by
\cref{eq:general_solution_two_populations}, with the initial
conditions \(a^{0, l}_{m, j} = \delta_{m0} \delta_{j0}\) and \(A^{0, l}_{m, j}
= 0\) where \(\delta_{ij}\) is the Kronecker delta.

It remains to derive a recurrence relation for the coefficients
\(a^{n, l}_{m, j}\) and \(A^{n, l}_{m, j}\).
Let us define the integrals \cref{eq:def-L,eq:def-V} so that
we can rewrite \cref{eq:iterative_two_populations} as
\cref{eq:P_recurrence_dev_1}.
The integrals \cref{eq:def-L,eq:def-V} have the solutions
respectively given by Eqs.~(\ref{eq:L_solution}) and (\ref{eq:V_solution}),
where the coefficients \(\xi_{m, j}^{q, r}\), \(\zeta_{m, j}^{q,
r}\), \(\Xi_{m, j}^{q, r}\), \(\Zeta_{m, j}^{q, r}\),
\(\eta_{m, j}^{q, r}\) and \(\theta_{m, j}^{q, r}\) also satisfy
recurrence relations which we give in the following
together with their initial conditions.

\begin{IEEEeqnarray}{rCl}
P_{n+1, l}(T) &=& \sum_{q=0}^{l} \int_0^{T-\tau} P_{n, q}(t)
\lambda \mathrm{e}^{-\lambda (T-t)} \frac{\left(\mu
\tau\right)^{l-q}}{(l-q)!} \mathrm{e}^{-\mu \tau} \, \mathrm{d} t
\nonumber\\
&& + \sum_{q=0}^{l} \int_{T-\tau}^T P_{n, q}(t) \lambda
\mathrm{e}^{-\lambda (T-t)}
\frac{\left(\mu (T-t)\right)^{l-q}}{(l-q)!}
\nonumber \\
&& \hphantom{+}\times \mathrm{e}^{-\mu (T-t)} \, \mathrm{d} t
\label{eq:iterative_two_populations}
\end{IEEEeqnarray}

\begin{IEEEeqnarray}{rCl}
P_{n, l}(t) &=& \mathrm{e}^{-\lambda t} \sum_{m=0}^{n}
\sum_{j=0}^{n+l} a^{n, l}_{m, j} \left( t - m \tau \right)^{j}_{+}
\nonumber \\
&& + \mathrm{e}^{-(\mu + \lambda) t} \sum_{m=0}^{n}
\sum_{j=0}^{n+l} A^{n, l}_{m, j} \left( t - m \tau \right)^{j}_{+}
\label{eq:general_solution_two_populations}
\end{IEEEeqnarray}

\begin{IEEEeqnarray}{rCl}
L_{m, j}^{q}(T) &=& \int_{T-\tau}^T \left( t - m \tau \right)^{j}_{+}
\frac{\left(\mu (T-t)\right)^{q}}{q!} e^{-\mu (T-t)} \,
\mathrm{d} t \label{eq:def-L}
\IEEEeqnarraynumspace\\
V_{m, j}^{q}(T) &=& \int_{T-\tau}^T \left( t - m \tau \right)^{j}_{+}
\frac{\left(\mu (T-t)\right)^{q}}{q!} \, \mathrm{d} t \label{eq:def-V}
\end{IEEEeqnarray}

\begin{widetext}


\begin{IEEEeqnarray}{rCl}
P_{n+1,l}(T) & = & \lambda e^{-\mu \tau} \sum_{q=0}^{l}
\frac{(\mu \tau)^q}{q!} \sum_{m=0}^{n} \sum_{j=0}^{n+l-q}
a^{n,l-q}_{m,j} \frac{(T - (m+1) \tau)^{j+1}_{+}}{j+1}  e^{-\lambda
T} + \lambda e^{-\mu \tau} \sum_{q=0}^{l}
\frac{(\mu \tau)^q}{q!} \sum_{m=0}^{n} \sum_{j=0}^{n+l-q}
A^{n,l-q}_{m,j} J_{m,j}(T)  e^{-\lambda T}
\nonumber\\ && + \lambda \sum_{q=0}^{l} \sum_{m=0}^{n}
\sum_{j=0}^{n+l-q} a^{n,l-q}_{m,j} L_{m,j}^{q}(T) e^{-\lambda T}
+ \lambda \sum_{q=0}^{l} \sum_{m=0}^{n}
\sum_{j=0}^{n+l-q} A^{n,l-q}_{m,j} V_{m,j}^{q}(T) e^{-(\mu + \lambda) T}
\label{eq:P_recurrence_dev_1}
\end{IEEEeqnarray}

\begin{IEEEeqnarray}{rCl}
L_{m, j}^{q}(T) &=& \sum_{r=0}^{j} \left( \xi_{m, j}^{q, r}
(T-m\tau)^r_{+} + \zeta_{m, j}^{q, r} (T-(m+1)\tau)^r_{+} \right)
\nonumber \\ &&+ \sum_{r=0}^{q}\left( \Xi_{m, j}^{q, r}
\mathrm{e}^{-\mu T}(T-m\tau)^r_{+}
+ \Zeta_{m, j}^{q, r} \mathrm{e}^{-\mu T} (T-(m+1)\tau)^r_{+} \right)
\label{eq:L_solution} \\ V_{m, j}^{q}(T) &=& \sum_{r=0}^{j+q+1}
\left( \eta_{m, j}^{q, r} (T-m\tau)^r_{+}
+ \theta_{m, j}^{q, r} (T-(m+1)\tau)^r_{+} \right) \label{eq:V_solution}
\end{IEEEeqnarray}

\end{widetext}

The initial conditions for the coefficients of Eq.~(\ref{eq:L_solution})
are given as follows. For \(j \geqslant 0\), if \( r \leqslant j \) then
\begin{align}
\xi_{m, j}^{0, r} & = - \frac{j!}{r!} \frac{1}{(-\mu)^{1+j-r}},
\\ \zeta_{m, j}^{0, r} &= e^{- \mu \tau} \frac{j!}{r!}
\frac{1}{(-\mu)^{1+j-r}},
\end{align}
if \( r \geqslant j+1 \) then
\begin{equation}
\xi_{m, j}^{0, r} = 0, \quad
\zeta_{m, j}^{0, r} = 0,
\end{equation}
for any \(r\),
\begin{align}
\Xi_{m, j}^{0, r} & = e^{\mu m \tau} \frac{j!}{(-\mu)^{1+j}} \delta_{r0},
\\ \Zeta_{m, j}^{0, r} &= - e^{\mu m \tau}
\frac{j!}{(-\mu)^{1+j}} \delta_{r0},
\end{align}
and for \(q \geqslant 1\),
\begin{align}
\xi_{m, 0}^{q, r} & = \frac{1}{\mu} \delta_{r0},
\\ \zeta_{m, 0}^{q, r} &= - \sum_{k=0}^{q} \frac{1}{k!} \mu^{k-1}
\tau^k e^{-\mu \tau} \delta_{r0},
\end{align}
if \( r \leqslant q \) then
\begin{align}
\Xi_{m, 0}^{q, r} & = -\frac{1}{r!} \mu^{r-1} e^{\mu m \tau},
\\ \Zeta_{m, 0}^{q, r} &= \sum_{k \geqslant r}^{q} \frac{1}{k!}
\mu^{k-1} \binom{k}{r} \tau^{k-r} e^{\mu m \tau},
\end{align}
if \( r \geqslant q+1 \) then
\begin{equation}
\Xi_{m, 0}^{q, r} = 0, \quad
\Zeta_{m, 0}^{q, r} = 0.
\end{equation}
The initial conditions for the coefficients of
Eq.~(\ref{eq:V_solution}) are given as follows: for \(j \geqslant 0\),
\begin{align}
\eta_{m, j}^{0, r} & = \frac{1}{j+1} \delta_{r,j+1},
\\ \theta_{m, j}^{0, r} &= - \frac{1}{j+1} \delta_{r,j+1},
\end{align}
for \(q \geqslant 1\),
\begin{align}
\eta_{m, 0}^{q, r}   & = \frac{\mu^q}{(q+1)!} \delta_{r,q+1},
\\
\theta_{m, 0}^{q, r} & = \frac{\mu^q}{(q+1)!} \tau^{q+1}
\delta_{r,0} - \frac{\mu^q}{(q+1)!} \binom{q+1}{r} \tau^{q+1-r},
\end{align}
and finally
\begin{equation}
\theta^{q, r}_{m, 0} = 0 \quad \text{for} \; r > q+1.
\end{equation}
We can derive recurrence relations for the coefficients using
integrations by parts, and the
results are given by
\cref{eq:sol-L-ksi-zeta,eq:sol-L-ksi-zeta-capital,eq:sol-V-eta-theta}.
Then \cref{eq:P_recurrence_dev_1} becomes
\cref{eq:P_recurrence_dev_2} and we finally obtain
recurrence relations for the coefficients \(a^{n, l}_{m, j}\)
and \(A^{n, l}_{m, j}\) in
\cref{eq:sol-a-0,eq:sol-a-j,eq:sol-A-j}.
We have thus obtained the analytical solution for the probability
\(P_{n, l}(T)\)
for the case where the dead time is a constant \(\tau\).

While it is straightforward to implement
\cref{eq:P_recurrence_dev_2} in a computer program
to compute the probabilities \(P_{n, l}(T)\), it is in fact
challenging to guarantee the
accuracy of the results. This is because the probability \(P_{n,
l}(T)\) is a number
between 0 and 1, and the coefficients in the intermediate steps can
be very large
due to the factorials in the numerators. Subtracting large numbers
to obtain a small
number is numerically unstable, hence we cannot fully trust the
numerical results.

\begin{align}
& \xi_{m, j+1}^{q+1, r} = \xi_{m, j+1}^{q, r} -\frac{j+1}{\mu}
\xi_{m, j}^{q+1, r}, \nonumber \\
& \zeta_{m, j+1}^{q+1, r} = \frac{(\mu \tau)^{q+1}}{(q+1)!}
\frac{ e^{-\mu \tau}}{-\mu} \delta_{r,j+1}
+ \zeta_{m, j+1}^{q, r} -\frac{j+1}{\mu} \zeta_{m, j}^{q+1, r},
\label{eq:sol-L-ksi-zeta}                                 \\
& \Xi_{m, j+1}^{q+1, r} = \Xi_{m, j+1}^{q, r} -\frac{j+1}{\mu}
\Xi_{m, j}^{q+1, r}, \nonumber \\
&\Zeta_{m, j+1}^{q+1, r} = \Zeta_{m, j+1}^{q, r}
-\frac{j+1}{\mu} \Zeta_{m, j}^{q+1, r},
\label{eq:sol-L-ksi-zeta-capital} \\
& \eta^{q+1, r}_{m, j} = \frac{\mu}{j+1} \eta^{q, r}_{m, j+1},  \nonumber \\
& \theta^{q+1, r}_{m, j} = - \frac{1}{j+1} \frac{(\mu \tau)^{q+1}}{(q+1)!}
\delta_{r,j+1} + \frac{\mu}{j+1} \theta^{q, r}_{m, j+1}.
\label{eq:sol-V-eta-theta}
\end{align}

\begin{widetext}
\begin{IEEEeqnarray}{rCl}
P_{n+1,l}(T) & = & \lambda e^{-\mu \tau} \sum_{q=0}^{l}
\frac{(\mu \tau)^q}{q!} \sum_{m=1}^{n+1} \sum_{j=1}^{n+l-q+1}
a^{n,l-q}_{m-1,j-1} \frac{(T - m \tau)^{j}_{+}}{j}  e^{-\lambda
T} \nonumber \\
&& + \lambda \sum_{q=0}^{l} \frac{(\mu \tau)^q}{q!}
\sum_{m=1}^{n+1} \sum_{k=0}^{n+l-q} A^{n,l-q}_{m-1,k}
\frac{k!}{\mu^{k+1}} e^{-\mu m \tau} (T-m\tau)^{0}_{+}
e^{-\lambda T} \nonumber \\
&& - \lambda \sum_{q=0}^{l} \frac{(\mu \tau)^q}{q!}
\sum_{m=1}^{n+1} \sum_{k=0}^{n+l-q} A^{n,l-q}_{m-1,k}
\sum_{j=0}^{k} \frac{k!}{j!} \frac{1}{\mu^{1+k-j}}
(T-m\tau)^{j}_{+} e^{-(\mu+\lambda) T} \nonumber \\
&& + \lambda \sum_{q=0}^{l} \sum_{m=0}^{n} \sum_{k=0}^{n+l-q}
a^{n,l-q}_{m,k} \sum_{j=0}^{k} \xi_{m,k}^{q,j}
(T-m\tau)^{j}_{+} e^{-\lambda T} \nonumber \\
&& + \lambda \sum_{q=0}^{l} \sum_{m=1}^{n+1} \sum_{k=0}^{n+l-q}
a^{n,l-q}_{m-1,k} \sum_{j=0}^{k} \zeta_{m-1,k}^{q,j}
(T-m\tau)^{j}_{+} e^{-\lambda T} \nonumber \\
&& + \lambda \sum_{q=0}^{l} \sum_{m=0}^{n} \sum_{k=0}^{n+l-q}
a^{n,l-q}_{m,k} \sum_{j=0}^{q} \Xi_{m,k}^{q,j}
(T-m\tau)^{j}_{+} e^{-(\mu + \lambda) T} \nonumber \\
&& + \lambda \sum_{q=0}^{l} \sum_{m=1}^{n+1} \sum_{k=0}^{n+l-q}
a^{n,l-q}_{m-1,k} \sum_{j=0}^{q} \Zeta_{m-1,k}^{q,j}
(T-m\tau)^{j}_{+} e^{-(\mu + \lambda) T} \nonumber \\
&& + \lambda \sum_{q=0}^{l} \sum_{m=0}^{n} \sum_{k=0}^{n+l-q}
A^{n,l-q}_{m,k} \sum_{j=0}^{k+q+1} \eta_{m,k}^{q,j}
(T-m\tau)^{j}_{+} e^{-(\mu + \lambda) T} \nonumber \\
&& + \lambda \sum_{q=0}^{l} \sum_{m=1}^{n+1} \sum_{k=0}^{n+l-q}
A^{n,l-q}_{m-1,k} \sum_{j=0}^{k+q+1} \theta_{m-1,k}^{q,j}
(T-m\tau)^{j}_{+} e^{-(\mu + \lambda) T}
\label{eq:P_recurrence_dev_2}
\end{IEEEeqnarray}

\begin{IEEEeqnarray}{rCl}
a^{n+1,l}_{m,0} & = & \lambda \sum_{q=0}^{l} \frac{(\mu
\tau)^{q}}{q!} \sum_{k=0}^{n+l-q} A^{n,l-q}_{m-1,k} \frac{k!}{\mu^{k+1}}
e^{-\mu m \tau}
+ \lambda \sum_{q=0}^{l} \sum_{k = 0}^{n+l-q} a^{n,l-q}_{m,k}
\xi^{q,0}_{m,k}
+ \lambda \sum_{q=0}^{l} \sum_{k = 0}^{n+l-q} a^{n,l-q}_{m-1,k}
\zeta^{q,0}_{m-1,k}
\label{eq:sol-a-0}
\end{IEEEeqnarray}
\begin{IEEEeqnarray}{rCl}
a^{n+1,l}_{m,j} & = & \lambda e^{-\mu \tau}
\sum_{q=0}^{\min{(l, n+l+1-j)}} \frac{(\mu \tau)^{q}}{q!}
\frac{a^{n,l-q}_{m-1,j-1}}{j}
+ \lambda \sum_{q=0}^{l} \sum_{k \geqslant j}^{n+l-q}
a^{n,l-q}_{m,k} \xi^{q,j}_{m,k}
+ \lambda \sum_{q=0}^{l} \sum_{k \geqslant j}^{n+l-q}
a^{n,l-q}_{m-1,k} \zeta^{q,j}_{m-1,k}
\label{eq:sol-a-j}
\end{IEEEeqnarray}

\begin{IEEEeqnarray}{rCl}
A^{n+1,l}_{m,j} & = & -\lambda \sum_{q=0}^{l} \frac{(\mu
\tau)^{q}}{q!} \sum_{k \geqslant j}^{n+l-q} A^{n,l-q}_{m-1,k}
\frac{k!}{j!} \frac{1}{\mu^{1+k-j}}
+ \lambda \sum_{q \geqslant j}^{l} \sum_{k = 0}^{n+l-q}
a^{n,l-q}_{m,k} \Xi^{q,j}_{m,k}
+ \lambda \sum_{q \geqslant j}^{l} \sum_{k = 0}^{n+l-q}
a^{n,l-q}_{m-1,k} \Zeta^{q,j}_{m-1,k} \nonumber \\
&& + \lambda \sum_{q=0}^{l} \sum_{k \geqslant
(j-q-1)_+}^{n+l-q} A^{n,l-q}_{m,k} \eta^{q,j}_{m,k}
+ \lambda \sum_{q=0}^{l} \sum_{k \geqslant (j-q-1)_+}^{n+l-q}
A^{n,l-q}_{m-1,k} \theta^{q,j}_{m-1,k}
\label{eq:sol-A-j}
\end{IEEEeqnarray}

\end{widetext}

\subsection{Exponential dead time (merged interval scenario)}
\label{app:exponential_ctmn_period}
The PDF for the dead time distribution is given by
\begin{equation}
q(t) = \frac{1}{\tau_\text{mean}} \mathrm{e}^{- t / \tau_\text{mean}} H(t).
\end{equation}
It follows that
\begin{equation}
p(1, T, \tau) = \delta(T - \tau) \frac{1}{\tau_\text{mean}}
\mathrm{e}^{-\lambda T} \lambda \mathrm{e}^{-\tau/\tau_{\text{mean}}},
\end{equation}
\begin{equation}
p(2, T, \tau) = \frac{1}{\tau_\text{mean}} \lambda \mathrm{e}^{-\lambda T}
\mathrm{e}^{-(T+\tau)/\tau_\text{mean}}.
\end{equation}
By replacing \(p(k, T, \tau)\) in the right hand side of
\cref{eq:recurrence_merged_scenario}
we obtain \cref{eq:merged_exponential_ctmn_period_recurrence_expansion}.
Again we have the same challenge as in the other cases, that is, we
need to subtract
large numbers to obtain a small value for the probability \(p(k, T,
\tau)\) and this
is not numerically reliable.

\begin{widetext}
\begin{IEEEeqnarray}{rCl}
&& p(k+1, T, \tau) \lambda^{-k} \tau_\text{mean}^{2-k}
\mathrm{e}^{\lambda T}
\nonumber \\ &=& \sum_{\substack{1 \leqslant j \leqslant k +
1\\1 \leqslant m \leqslant k}}
\left(\frac{T}{\tau_\text{mean}}\right)^{j}
\mathrm{e}^{- \frac{T m}{\tau_\text{mean}}}
\sum_{\mathclap{0 \leqslant g \leqslant j - 1}}
\frac{{F}_{j - 1,g,m,m}}{j - g}
- \sum_{\substack{1 \leqslant j \leqslant k + 1\\1 \leqslant m
\leqslant k}}
\left(\frac{T}{\tau_\text{mean}}\right)^{j}
\mathrm{e}^{- \frac{T m + \tau}{\tau_\text{mean}}}
\sum_{\mathclap{0 \leqslant g \leqslant j - 1}}
\frac{{F}_{j - 1,g,m,m}}{j - g}
\nonumber \\ && +
\sum_{{\substack{0 \leqslant l \leqslant j - 1\\1 \leqslant j
    \leqslant k + 1\\
1 \leqslant m \leqslant k}}} \frac{T^{l} \tau^{j - l}
\tau_\text{mean}^{- j}
\mathrm{e}^{\frac{- T m - \tau}{\tau_\text{mean}}} {F}_{j -
1,l,m,m}}{j - l}
- \sum_{\mathclap{\substack{0 \leqslant l \leqslant j - 1\\1 \leqslant j
    \leqslant k + 1\\
1 \leqslant m \leqslant k}}} \frac{T^{l} \tau^{j - l}
\tau_\text{mean}^{- j}
\mathrm{e}^{\frac{- T m}{\tau_\text{mean}}} {F}_{j - 1,l,m,m}}{j - l}
\nonumber \\ && +
\sum_{\substack{0 \leqslant j \leqslant k\\1 \leqslant m \leqslant k}}
\left(\frac{T}{\tau_\text{mean}}\right)^{j} \mathrm{e}^{-
\frac{T m}{\tau_\text{mean}}}
\sum_{\mathclap{\substack{0 \leqslant g \leqslant j\\j
\leqslant d \leqslant k\\1 \leqslant h \leqslant k, h \neq m}}}
\frac{\left(-1\right)^{d - j}
\left(d - g\right)! {F}_{d,g,h,m}}{\left(j - g\right)! \left(h
- m\right)^{d+1-j}}
\nonumber \\ && +
\sum_{\substack{0 \leqslant j \leqslant k\\1 \leqslant m \leqslant k}}
\left(\frac{T}{\tau_\text{mean}}\right)^{j} \mathrm{e}^{-
\frac{T m + \tau}{\tau_\text{mean}}}
\sum_{\mathclap{\substack{0 \leqslant g \leqslant j\\j
\leqslant d \leqslant k\\1 \leqslant h \leqslant k, h \neq m}}}
\frac{\left(-1\right)^{d - j+1}
\left(d - g\right)! {F}_{d,g,h,m}}{\left(j - g\right)! \left(h
- m\right)^{d+1-j}}
\nonumber \\ && +
\sum_{\substack{0 \leqslant l \leqslant j\\
1 \leqslant j \leqslant k + 1\\1 \leqslant m \leqslant k}}
T^{l} \tau^{j - l} \tau_\text{mean}^{- j}
\mathrm{e}^{\frac{- T m - \tau (1-m)}{\tau_\text{mean}}}
\sum_{\mathclap{\substack{0 \leqslant w \leqslant i\\
    0 \leqslant i \leqslant \min{(g, j-1)}\\
0 \leqslant g \leqslant d\\ j-1 \leqslant d \leqslant k}}}
\frac{\left(-1\right)^{j - l + w}
{\binom{g}{i}} {\binom{i}{w}} {\binom{j}{l}} \left(d - i\right)!
{F}_{d,g,m,m}}{\left(- i + j + w\right) m^{d - j + 2} \left(- i
+ j - 1\right)!}
\nonumber \\ && +
\sum_{\substack{0 \leqslant l \leqslant j
\\1 \leqslant j \leqslant k + 1\\1 \leqslant m \leqslant k}}
T^{l} \tau^{j - l} \tau_\text{mean}^{- j} \mathrm{e}^{\frac{- T
m - \tau}{\tau_\text{mean}}}
\sum_{\mathclap{\substack{(l-u-1)_+ \leqslant w \leqslant j - u - 1
    \\0 \leqslant i \leqslant u\\0 \leqslant u \leqslant \min{(g, j-1)}
\\0 \leqslant g \leqslant d\\j-1 \leqslant d \leqslant k}}}
\frac{\left(-1\right)^{i - l + u + w}
{\binom{g}{u}} {\binom{u}{i}} {\binom{j - u - 1}{w}}
{\binom{u + w + 1}{l}} \left(d - u\right)!
{F}_{d,g,m,m}}{\left(i + w + 1\right)
m^{d - j + 2} \left(j - u - 1\right)!}
\nonumber \\ && +
\sum_{\substack{0 \leqslant l \leqslant j\\0 \leqslant j \leqslant k
\\1 \leqslant m \leqslant k}}
T^{l} \tau^{j - l} \tau_\text{mean}^{- j}
\mathrm{e}^{\frac{-T m - \tau (1-m)}{\tau_\text{mean}}}
\sum_{\mathclap{\substack{(u-j)_+ \leqslant i \leqslant u\\ (j-u)_+
    \leqslant w \leqslant d - u
    \\0 \leqslant u \leqslant g\\0 \leqslant g \leqslant d\\j
    \leqslant d \leqslant k
\\1 \leqslant h \leqslant k, h \neq m}}}\;\;
\frac{\left(-1\right)^{i - l + u + w}
{\binom{g}{u}} {\binom{j}{l}} {\binom{u}{i}} \left(d - u\right)!
\left(i + w\right)! {F}_{d,g,h,m}}{m^{d - u - w + 1} \left(h -
m\right)^{u+w-j+1} w! \left(i + j - u\right)!}
\nonumber \\ && +
\sum_{\substack{0 \leqslant l \leqslant j\\0 \leqslant j
\leqslant k\\1 \leqslant m \leqslant k}}
T^{l} \tau^{j - l} \tau_\text{mean}^{- j}
\mathrm{e}^{\frac{- T m - \tau}{\tau_\text{mean}}}
\sum_{\mathclap{\substack{0 \leqslant i \leqslant \min{(v, j-l, j-w)}\\0
    \leqslant w \leqslant \min{(u, j)}
    \\ (j-u)_+ \leqslant v \leqslant d - u
    \\0 \leqslant u \leqslant g\\0 \leqslant g \leqslant d\\j
    \leqslant d \leqslant k\\
1 \leqslant h \leqslant k, h \neq m}}}
\frac{\left(-1\right)^{v - i - l - w - 1}
{\binom{g}{u}} {\binom{u}{w}}
{\binom{v}{i}} {\binom{j - i}{l}} \left(d - u\right)! \left(v
- i + u - w\right)!
{F}_{d,g,h,m}}{ m^{d - u - v + 1} \left(h - m\right)^{u+v-j+1}
v! \left(j - w - i\right)!}
\nonumber \\ && +
\sum_{\substack{0 \leqslant l \leqslant j\\0 \leqslant j \leqslant k\\
1 \leqslant m \leqslant k\\2 \leqslant q \leqslant k + 1 \\ q \neq m + 1}}
T^{l} \tau^{j - l} \tau_\text{mean}^{- j} \mathrm{e}^{\frac{- T
m - \tau (q-m)}{\tau_\text{mean}}}
\sum_{\substack{j \leqslant d \leqslant k}}
\frac{\left(-1\right)^{d - j} \left(d - l\right)!
{F}_{d,l,m,q - 1}}{\left(j - l\right)! \left(m - q + 1\right)^{d + 1 - j}}
\nonumber \\ && +
\sum_{\substack{0 \leqslant l \leqslant j\\0 \leqslant j \leqslant k\\
1 \leqslant m \leqslant k\\1 \leqslant q \leqslant k \\ q \neq m}}
T^{l} \tau^{j - l} \tau_\text{mean}^{- j} \mathrm{e}^{\frac{- T
m - \tau (q-m)}{\tau_\text{mean}}}
\sum_{\substack{j \leqslant d \leqslant k}}
\frac{\left(-1\right)^{d - j + 1} \left(d - l\right)!
{F}_{d,l,m,q}}{\left(j - l\right)! \left(m - q\right)^{d + 1 - j}}
\label{eq:merged_exponential_ctmn_period_recurrence_expansion}
\end{IEEEeqnarray}
\end{widetext}

\end{document}